\documentclass[draft]{agujournal2019}
\usepackage{url} 
\usepackage{lineno}
\usepackage[inline]{trackchanges} 
\usepackage{soul}
\usepackage{amsmath,amssymb }
\usepackage{mathtools}   
\usepackage{empheq}      
\usepackage{enumitem}

\newcommand{\be}{\begin{equation}}
\newcommand{\ee}{\end{equation}} 

\newcommand{\bu}{{\bf u}}

\newcommand{\bx}{{\bf x}}
\newcommand{\by}{{\bf y}}

\newcommand{\lb}{\label}

\newcommand{\cS}{{\mathcal S}}

\newcommand{\cD}{{\mathcal D}}
\newcommand{\cB}{{\mathcal B}}
\newcommand{\cC}{{\mathcal C}}
\newcommand{\bzero}{{\textbf{0}}}

\newcommand{\Bdr}{{\partial\cD}}
\newcommand{\tBdr}{{\partial\widetilde{\cD}}}
\newcommand{\ext}{{\mathrm{ext}}}
\newcommand{\fil}{\mathrm{fill}}

\newcommand{\grad}{{\mbox{\boldmath $\nabla$}}}

\newcommand{\bdot}{{\mbox{\boldmath $\cdot$}}}

\newcommand{\paren}[1]{\left( #1 \right)}

\newcommand{\absv}[1]{\left| #1 \right|}

\usepackage{xcolor}

\draftfalse

\begin{document}

%
%


\title{
Harmonic Extension for Multiscale Analysis and Modeling Near Boundaries, with an Ocean Application
}

%
%




\authors{Benjamin A. Storer\affil{1}, Mehrnoush Kharghani\affil{1}, Alistair Adcroft \affil{2}, Hussein Aluie\affil{1,3,4}}

\affiliation{1}{Department of Mechanical Engineering, University of Rochester, Rochester, NY, USA}
\affiliation{2}{Program in Atmospheric and Oceanic Sciences, Princeton University, Princeton, NJ, USA}
\affiliation{3}{Laboratory for Laser Energetics University of Rochester, Rochester, NY, USA}
\affiliation{4}{Department of Mathematics, University of Rochester, Rochester, NY, USA}




\correspondingauthor{Hussein Aluie}{hussein@rochester.edu}



\begin{keypoints}
\item Harmonic extension: principled method to fill irregular land regions by solving Laplace problem consistent with ocean boundary conditions
\item The approach guarantees a unique extension with minimal spatial variability  that satisfies the boundary conditions
\item The method is easy to implement numerically for use in multiscale analysis and modeling, including in multiphase flows
\end{keypoints}

%
%

%
%


\begin{abstract}
Treatment of fields near domain boundaries is a long-standing problem in signal processing that has come into renewed focus following recent efforts in convolution-based multiscale coarse-graining and in machine-learned parameterizations due to ocean boundary artifacts. Here, we propose a general and principled method for extending fields beyond the domain boundaries by solving a Laplace boundary-value problem. Construction of the harmonic extension is well-posed, including uniqueness, and is consistent with the boundary conditions by design. The formulation applies to irregular boundaries such as discretized coastlines. The harmonic extension is physically desirable since it has minimum spatial variability among all admissible extensions satisfying the boundary conditions. The method is simple to implement using well-established numerical approaches,  and is broadly applicable to extending oceanic variables over land boundaries. Other applications include machine learning parametrization and subgrid modeling of wall-bounded flows and multiphase flows. We demonstrate the method by extending sea-surface temperature (SST) over land using fixed temperature (Dirichlet) and no-flux (Neumann) boundary conditions: the land-filled solution is smooth with SST values between the coastal minimum and maximum.

\end{abstract}

\section*{Plain Language Summary}
Earth system models and  datasets from satellite observations often only define ocean variables where there is water. Yet, many modern analysis tools and machine learning methods require information over land regions, beyond the ocean boundary. To make the data ``workable,'' land points have been filled using simple tricks such as copying nearby values, setting them to zero, or heavily smoothing the field. These ad hoc fixes can distort the real ocean signal, especially near coasts where many important processes occur. Here, we introduce a mathematically well-posed way to fill land points by regarding the ocean field as an undulated/stretched membrane due to the vigorous and complex circulation. Along the coastline we keep the ocean values fixed in a manner consistent with the data, while over land we let the membrane relax to its smoothest possible shape. This is equivalent to solving a classical Laplace equation, which guarantees a unique, minimally variable extension that is fully consistent with the coastal values, \textit{i.e.} boundary conditions. Our approach is easy to implement with standard numerical tools and works for complex coastlines, vector fields, and curved surfaces such as the globe.

%
%

%


%
%
%
%

\section{Introduction and Motivation}
Classical signal processing has long relied on domain ``extensions'' to enable the use of Fourier methods on non-periodic data. A standard strategy is to embed a finite signal into a larger periodic domain so that the fast Fourier transform (FFT) can be applied efficiently, typically through simple devices such as periodic extension, windowing, mirroring, or zero padding at the boundaries \cite{boyd2001chebyshev,boyd2002comparison}. These constructions offer practical solutions for handling non-periodic data with FFTs, but they do not in general respect the boundary conditions implied by the underlying dynamics, which are seldom periodic.



Another scale-analysis method to probe multiscale processes relies on convolution-based coarse-graining \cite{Leonard74,germano1992turbulence,Meneveau1994,Eyink95,EyinkAluie09}. This approach resolves processes at different scales in space-time, and its generalization to spherical surfaces \cite{aluie2019conv} 
has led to its rapid adoption in geophysical applications \cite{aluie2018mapping,srinivasan2019submesoscale,schubert2020submesoscale,grooms2021diffusion,rai2021scale,steinberg2022seasonality,schubert2023open,juricke2023scale,solodoch2023basin,loose2023comparing,li2024eddy,xue2024surface,liu2024spatial,barkan2024eddy,kouhen2024convective,danilov2024extracting,yu2024intensification,shaham2025spectral,rai2025atmospheric}. 
 To probe processes at length scale $\ell$ on the globe, the method relies on low-pass spatial filtering, constructed so that the convolution commutes with spatial derivatives on spherical surfaces and thus preserves scale-dependent symmetries \cite<see>[]{aluie2019conv}. Filtering near boundaries involves a weighted average over a ball of radius $\ell/2$ around $\bx$, which may cross the domain boundary. For oceanic quantities such as velocity or tracers, the kernel may overlap land points. Motivated by oceanic applications, we refer to regions outside the domain boundaries as ``land,'' while keeping in mind that the applications are more general. One option is to deform the filtering kernel to avoid land, but then convolutions with such a ``deformed kernel'' do not commute with spatial derivatives \cite{aluie2019conv}. For a comparison between ``deformed'' and ``fixed'' kernels, we refer the reader to \citeA{grooms2021diffusion,buzzicotti2022coarse}. 
 
 In past studies \cite{zhao2018inviscid,aluie2018mapping}, it was argued that a physically sound choice for land values is one that is consistent with the boundary conditions (BCs) governing the dynamics. For the velocity field, past oceanographic works \cite{storer2022global,storer2023global,khatri2024scale} had chosen to treat land as zero-velocity water, motivated by the fact that most general circulation models impose rigid-wall, no-slip BCs for velocity. For scalar fields beyond the top and bottom boundaries in canonical Rayleigh-Taylor- (buoyancy-) driven flows, \citeA{zhao2018inviscid,zhao2022scale} kept density constant (zero normal gradient) and extended pressure according to the hydrostatic relation $\partial_z p=-\rho g$ (non-homogeneous Neumann BC). However, for more general scenarios, such as extending oceanic tracers over land, there remains an infinite family of possible extensions that all satisfy the boundary conditions.

 A physically consistent strategy for land treatment (or domain extension) is essential not only for coarse-graining, but also for traditional signal-processing tools such as Fourier-based methods. For Earth system data, spherical harmonics provide a natural global basis defined over the entire sphere, but they are not restricted to the ocean domain, which is one reason they have been seldom used in oceanic applications. During the early days of satellite altimetry, there were attempts to utilize spherical harmonics \cite{wunsch1991global,wunsch1995global}; however, land values were treated without dynamical justification, for example by setting them at each latitude to the zonal average \cite{wunsch1995global}, with the acknowledgment that ``\dots no claim that we have made the best possible choice.'' It appears that the use of spherical harmonics for ocean analysis was largely abandoned after these early studies. More recently, the results of \citeA{buzzicotti2023spatio} have provided  justification for using spherical harmonics on the global ocean, with convolution-based coarse-graining offering a guide for treating land in a manner consistent with the governing boundary conditions.


Treatment of points beyond the domain boundaries is also a problem in machine-learned parameterizations, as highlighted recently by \citeA{zhang2025addressing} in an oceanographic context. Convolutional neural network (CNN) parameterizations of subgrid physics perform well in the open ocean but often degrade near coastlines, where sliding convolution stencils intersect land and thus encounter inputs that are out-of-sample relative to the training data. This boundary contamination propagates across layers in multi-layer CNNs, producing coastal artifacts (e.g., spurious shear and eddies), inflated errors, and in some cases numerical instability. Two practical remedies were proposed in \citeA{zhang2025addressing}, namely zero padding and replicate padding (a quasi-Neumann boundary condition) beyond the boundary. In particular, replicate padding was found to help minimize the propagation of extreme values that can contaminate model fields or cause simulations to fail.

\begin{figure}
    \centering
    \includegraphics[width=0.99\linewidth]{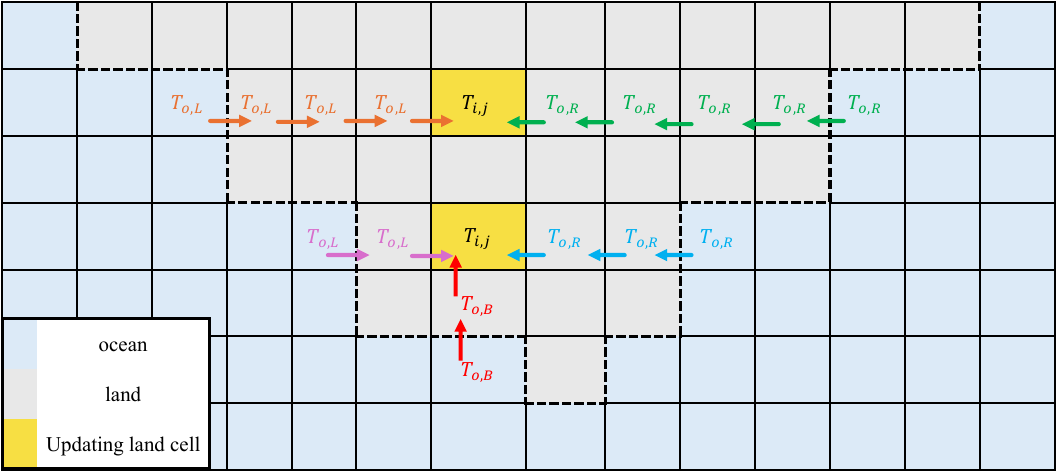}
    \caption{\textbf{Schematic illustration of replicate-padding.}
    Schematic illustrating why replicate padding does not yield a unique solution for filling interior land regions (grey). Each yellow cell represents a land point to be filled, while arrows indicate replicate-padding of data from neighboring ocean points (blue). How far inland one needs to fill depends on the lengthscales $\ell$ being parameterized, with larger scales requiring deeper padding over land, up to distances of order $\ell/2$ from the boundary. 
    Because these land cells receive conflicting values from different directions, replicate padding introduces ambiguity and non-uniqueness in the filled field.}
    \label{fig:Schematic}
\end{figure}

However, two issues remain: (1) consistency with the boundary conditions of the governing dynamics, and (2) lack of uniqueness, in that imposing a homogeneous Neumann (no-flux) condition does not uniquely determine land values at arbitrary distances from the boundary. How far one needs to fill into land depends on the length scales being parameterized, with larger scales requiring deeper padding over land, up to distances of order $\sim \ell/2$ from the boundary. The non-uniqueness is most easily seen around an island (or peninsula), as in Figure~\ref{fig:Schematic}, where the replicate-padding strategy is unlikely to yield a self-consistent value at interior land points that is compatible with the entire surrounding boundary.

In this study, we propose a general framework for physically consistent land-filling of fields defined on bounded domains. Our method is based on solving Laplace’s equation with boundary conditions that respect the physical constraints at the land-ocean interface. We illustrate the approach using sea-surface temperature (SST), but the harmonic extension method can be useful for a broad range of applications that require domain extension, including Fourier-based methods and local spatial filtering in Large Eddy Simulation (LES) of wall-bounded flows.

    \section{Method}\label{GE}

\subsection{Harmonic extension beyond the boundary}
Fix a time $t$ and consider a field $u(\bx)$ defined over a domain
$\cD$ (e.g. the global ocean), subject to boundary conditions on
$\Bdr$ that are imposed by the governing dynamics. Let $\widetilde{\cD}$ be a larger \emph{embedding domain} within which $\cD$ and $\Bdr$ are subsets, $(\cD\cup \Bdr) \subset \widetilde{\cD}$. For the ocean, $\widetilde{\cD}$ can be the entire spherical surface $\cS_r$ of radius $r$. More generally, for wall-bounded flows, $\widetilde{\cD}$ can be the $n$-dimensional real space $\mathbb{R}^n$. For the \emph{fill region} (e.g.\ land points beyond the coastline), we write
\[
\cD_{\fil} := \widetilde{\cD} \setminus (\cD\, \cup \Bdr),\]
\textit{i.e.,} $\cD_{\fil}$ consists of all points in $\widetilde{\cD}$ that are neither in $\cD$ nor in $\Bdr$. The fill region $\cD_{\fil}$ is bounded by $\Bdr$, which forms the interface with the original domain, and possibly by an outer boundary, $\partial\widetilde{\cD}$, of the embedding domain. If the embedding domain $\widetilde{\cD}$ has no boundaries, such as when it is the entire spherical surface $\cS_r$ or real space $\mathbb{R}^n$, then $\partial\cD$ is the only boundary for the fill region $\cD_{\fil}$. 

Given $u(\bx)$ on $\cD$, we define an extended field $u_\ext(\bx)$ as follows. First, we solve a Laplace boundary value problem in the fill region:
\begin{subequations}\label{eq:HarmonicExtension}
\begin{empheq}[left=\empheqlbrace]{align}
&\nabla^2 u_*(\bx) = 0, & \bx \in \cD_{\fil}, \label{eq:HarmonicExtension-a}\\
&\cB u_*(\bx) = \cB u(\bx), & \bx \in \Bdr, \label{eq:HarmonicExtension-b}\\
&\cC u_*(\bx) = g(\bx), & \bx \in \tBdr, \label{eq:HarmonicExtension-c}
\end{empheq}
\end{subequations}
where $\nabla^2$ is the Laplace operator or, more generally, the Laplace-Beltrami operator on curved surfaces such as the sphere. $\cB $ denotes the boundary operator that enforces consistency with the original field, e.g.\ Dirichlet: $\cB u = u$ so that $u_*|_{\Bdr}=u|_{\Bdr}$, or Neumann: $\cB u=\partial_n u$ so that $\partial_n u_*|_{\Bdr}=\partial_n u|_{\Bdr}$ to match normal flux, where $\partial_n$ denotes the outward normal derivative from the perspective of the original domain $\cD$. In eq.~\eqref{eq:HarmonicExtension-c}, $\cC$ encodes auxiliary conditions on the outer boundary of the embedding domain, e.g.\ homogeneous Neumann $\partial_n u_*|_{\tBdr} = 0$ or homogeneous Dirichlet $u_*|_{\tBdr}=0$. If $\widetilde{\cD}$ is compact with no boundaries, such as in the case of the spherical surface, eq.~\eqref{eq:HarmonicExtension-c} is irrelevant since $\tBdr$ does not exist. A solution $u_*$ that satisfies the Laplace equation~\eqref{eq:HarmonicExtension} is called a harmonic function \cite{axler2013harmonic}.

The \emph{harmonic extension} of $u$ to $\widetilde{\cD}$ is then defined as
\begin{equation}
  u_{\ext}(\bx) :=
  \begin{cases}
    u(\bx), & \bx \in \cD\cup\Bdr,\\[4pt]
    u_*(\bx), & \bx \in \cD_{\fil},
  \end{cases}
\end{equation}
where $u_*$ is the unique solution for eq.~\eqref{eq:HarmonicExtension}. In what follows, without loss of generality, we assume that $\Bdr$ is the only boundary for $\cD_\fil$. 

Note that $\cD_\fil$ can be multiply connected, for example when the land to be filled completely surrounds an inland sea such as the Caspian Sea. The method can also be easily applied to flows with domain boundaries that deform or evolve in time. Examples include ocean-ice boundaries and multiphase flows, such as sea-spray droplets and gas bubbles at the air-sea interface \cite{deike2022mass}. For such applications with time-varying boundaries, the Laplace boundary value problem~\eqref{eq:HarmonicExtension}  has to be solved for each time and with an updated boundary interface $\Bdr$.

\subsection{Physical Justification}\label{sec:PhysicsJustify}
The Laplace boundary value problem~\eqref{eq:HarmonicExtension} is well-posed following basic theory of partial differential equations \cite<e.g.,>{evans2022partial}. Well-posedness means \cite{hadamard1902problemes} that solution $u_*$ is guaranteed to exist, is unique, and that small changes in the boundary conditions yield proportionally small changes in the solution (\textit{i.e.}, stability). This makes the method appealing from a physical perspective and also from considerations of its numerical implementation.

Moreover, the harmonic extension is physically desirable since $u_*$ is the field with minimum spatial variability while satisfying the boundary conditions. More precisely, $u_*$ is the minimizer of the Dirichlet energy \cite{evans2022partial}, $E(v)$, in the fill region,
\begin{equation}
  u_* = \arg\min\left\{E(v):=
    \int_{\cD_{\fil}} |\nabla v(\bx)|^2\,\mathrm{d}\bx
    \,\middle|\,
    v \text{ satisfies the BCs in \eqref{eq:HarmonicExtension}}
  \right\}.
\end{equation}
The physical reason we desire $u_*$ to have the least possible spatial variability is that the fill region is ``quiescent'', without sources of variability beyond that which is imposed at the domain boundaries.
In other words, $u_*$ serves as a background field of minimum variability. This is not necessarily the case of non-linear interpolation (or extrapolation) schemes, such as a cubic interpolant.

Related to minimizing the Dirichlet energy, the harmonic solution $u_*$ of eq.~\eqref{eq:HarmonicExtension} satisfies the \emph{maximum principle}: extrema values of $u_*$ lie at the boundary and cannot occur in the interior of the fill region, $\cD_\fil$. In fact, the value of $u_*(\bx)$ at every interior point $\bx\in\cD_\fil$ is a weighted average of the boundary values. This can be easily seen from the Green's function representation of the solution \cite{gilbarg1977elliptic}. For example, in the case of Dirichlet boundary conditions at the interface, $u_*|_{\Bdr}=u|_{\Bdr}$ in eq.~\eqref{eq:HarmonicExtension-b},
\be
u_*(\bx) = \int_{\by\in\partial\cD} P(\bx,\by)\, u(\by)\, \mathrm{d}\by,
\hspace{1cm} 
P(\bx,\by) = -\partial_{n_\by} G(\bx,\by).
\ee
Here, $G(\bx,\by)$ is the Laplacian Green's function ($G(\bx,\by)=-(2\pi)^{-1}\log(|\bx-\by|)$ in two-dimensions\footnote{On the spherical surface, $G(\bx,\by)=-(2\pi)^{-1}\log(\sin{\frac{\gamma}{2}})$, where $\gamma\in[0,\pi]$ is the geodesic angle subtended by the geodesic arc connecting $\bx$ and $\by$, which reduces to the flat 2D expression for small angles.}) and $\partial_{n_\by}$ denotes the  normal derivative at $\by\in\partial\cD$. This makes explicit that interior values of $u_*(\bx)$ are weighted averages of boundary values, with weight $P(\bx,\cdot)$ known as the Poisson kernel \cite{axler2013harmonic}. Any harmonic solution is infinitely smooth (real-analytic) in the interior of $\cD_\fil$, independent of boundary smoothness.

Putting the preceding discussion in a wider context, the Laplacian acts as a local high-pass filter: its sign diagnoses whether the value of a field $u_*(\bx)$ at location $\bx$ lies above or below its neighborhood average \cite{gilbarg1977elliptic}. If $\nabla^2 u_*(x)\ge 0$, then $u_*$ is, locally, below its neighborhood average (a negative anomaly); if $\nabla^2 u_*(x)\le 0$, it is above its neighborhood average (a positive anomaly). A function with $\nabla^2 u_*\ge 0$ throughout the domain is called \emph{subharmonic}; in one dimension this coincides with convexity, e.g. $u_*(x)=x^2$. A function with $\nabla^2 u_*\le 0$ everywhere is \emph{superharmonic}. A harmonic function satisfies $\nabla^2 u_*=0$ and obeys the mean-value property: its value at each point equals the average of its surroundings within a ball of \emph{any radius} contained in the domain. In effect, harmonic fields are very large-scale fields, the largest possible while satisfying the boundary conditions.

\subsection{Technical Considerations}\label{sec:MathSubtle}
In the case of Neumann BCs, 
\be
\begin{cases}
\nabla^2 u_* = 0 & \text{in }\cD_\fil,\\
\partial_n u_* = g & \text{on }\partial\cD,
\end{cases}
\ee
existence of a solution requires the solvability condition
  \[
  \int_{\partial\cD} g(\bx)\,\mathrm{d}\bx = 0
  \]
on the imposed flux $g$. In other words, the net (integrated) flux along the boundary has to vanish. Moreover, $u_*$ is unique up to an additive constant in the pure Neumann case.

Another consideration pertains to smoothness of the interface boundary $\Bdr$. Whether $u_*$ attains $u|_\Bdr$ pointwise everywhere along the boundary $\Bdr$ depends on the boundary regularity at each point \cite<e.g.,>{gilbarg1977elliptic}. 
Polygonal/polyhedral ocean domains $\cD$ (e.g., with corners arising from discretized coastlines) are sufficiently regular (Lipschitz) to render the Laplace boundary value problem~\eqref{eq:HarmonicExtension} well-posed in the usual weak (Sobolev) sense. In particular, there exists a unique locally-averaged (e.g., cell-averaged) solution $u_*$ such that both $u_*$ and its gradient are square-integrable, and $u_*$ has the prescribed values on $\Bdr$. This weak solution is precisely the object approximated by standard numerical schemes such as finite volume discretizations. For general bounded domains with highly irregular boundaries, such as cusps or fractal boundaries, the harmonic solution with boundary data $u|_\Bdr$ attains these boundary values at all regular points, but may fail to do so at irregular points \cite{gilbarg1977elliptic}. In practice, however, any fixed numerical discretization (or grid resolution) replaces the interface $\Bdr$ by a polygonal approximation, for which the Laplace problem is again well-posed. In other words, for each fixed grid representation of the interface $\Bdr$, solving the Laplace boundary value problem~\eqref{eq:HarmonicExtension} is well-posed both as a continuum problem and as a discrete numerical problem. Potential issues associated with rough boundaries arise only when considering the convergence of $u_*$ as the grid is refined and the discrete boundary tends to its irregular limit set.

\subsection{Generalizations}\lb{sec:Generalizations}


Although we have described the harmonic extension of a scalar field in eq.~\eqref{eq:HarmonicExtension}, the method can be naturally generalized to vector fields and higher rank tensors. If $\cD_\fil$ is a flat (Euclidean) space, then this is accomplished easily by solving problem~\eqref{eq:HarmonicExtension} for each of the vector's Cartesian components separately. This yields a harmonic vector- (or \mbox{tensor-)} extension over $\cD_\fil$ that satisfies the imposed boundary conditions at $\Bdr$. If $\cD_\fil$ is not flat, such as for oceanic applications, then one needs to replace the scalar Laplacian in eq.~\eqref{eq:HarmonicExtension-a} with the vector Laplacian (the covariant or Bochner Laplacian) that is intrinsic to the manifold \cite<e.g.,>[]{gilbarg1977elliptic}. While it is possible in curved space to still perform the harmonic extension component-wise using the scalar Laplacian, harmonicity would then be satisfied only for the vector components in that particular coordinate basis and would not yield a vector field that is harmonic in the geometric sense. On curved surfaces, the vector Laplacian is intrinsically different from the scalar Laplacian: it mixes vector components and involves additional metric-dependent terms arising from curvature. In Euclidean (flat) settings with Cartesian components, the vector Laplacian reduces to the component-wise scalar Laplacian since there is no curvature.

The harmonic extension method accommodates a wide range of boundary conditions for the velocity, including more general cases beyond the standard no-slip and free-slip formulations, for example those arising in subgrid wall-models \cite{bose2018wall}. If the vector field carries additional structure (e.g.\ an incompressible velocity field with $\nabla\cdot\mathbf{\bu}=0$), one may instead wish to enforce such constraints in the extension. In that case the extension is defined by a vector-valued elliptic system, for example a Stokes-type problem
\begin{subequations}\label{eq:Stokes}
\begin{empheq}[left=\empheqlbrace]{align}
&\nabla^2\bu_*(\bx) - \grad p(\bx) = 0, &  \bx\in\cD_\fil,\\
&\grad\cdot\bu_*(\bx) = 0, & \bx\in\cD_\fil,\\
&    \cB\bu_*(\bx) = \cB\bu(\bx), & \bx\in\Bdr,
\end{empheq}
\end{subequations}
with appropriate boundary conditions on $\Bdr$. Here, $p(\bx)$ is analogous to pressure and serves the role of a Lagrange multiplier to enforce the divergence-free condition. Thus, the idea of harmonic extension can be adapted to
structured vector fields by replacing the scalar Laplace equation with an
appropriate vector elliptic system. In theoretical physics, harmonic extensions have been used as background reference fields relative to which the degree of knottedness (helicity) of vorticity and magnetic field lines can be quantified in arbitrary material regions co-moving with the flow \cite{berger1984topological,FinnAntonsen1985,aluie2017coarse,soltani2023galilean}.

\subsection{Special Cases}\label{SpecialCases}
Past oceanographic studies \cite{storer2022global,storer2023global,khatri2024scale} chose to treat land as zero velocity water. The choice of filling land with $\bu_*=\bzero$ is a special case of the harmonic extension method. Indeed, for an oceanic circulation with no-slip and rigid wall Dirichlet BCs, $\cB \bu(\bx)=\bu(\bx)=\bzero$ at $\Bdr$, commonly used in general circulation models, the Laplace boundary value problem \eqref{eq:HarmonicExtension} has the homogeneous BC $\bu_*(\bx) = \bzero$ at $\Bdr$ in eq.~\eqref{eq:HarmonicExtension-b}. For an embedding domain $\widetilde{\cD}$ that is the entire globe, $\tBdr$ does not exist and eq.~\eqref{eq:HarmonicExtension-c} is irrelevant. The solution to a Laplace boundary value problem with Dirichlet boundary conditions set to zero yields the solution $\bu_*=\bzero$ over $\cD_\fil$ by the maximum principle: the minimum and maximum of $\bu_*$ should lie at $\Bdr$, but $\bu_*=\bzero$ on $\Bdr$. Therefore, $\bu_*=\bzero$ over $\cD_\fil$, regardless of curvature considerations\footnote{For rigid-wall no-slip BCs on the ocean velocity $\bu$, $\int_{\cD_\fil}\mathrm{d}\bx\, \bu_*\bdot \nabla^2_{\text{vec}}\bu_* = \int_{\cD_\fil}\mathrm{d}\bx\, |\grad\bu_*|^2 =0$ for the harmonic vector field $\bu_*$ over land. Therefore, the field is constant with $\bu_*=\bzero$ to match the BCs.} mentioned in section~\ref{sec:Generalizations}.

Another special case is the extension of pressure and density fields in 
a canonical Rayleigh-Taylor instability (buoyancy-) driven flow \cite{Rayleigh83, Taylor50}. The initial setup of the canonical problem has a heavy fluid on top of a light fluid in the presence of gravity $-g \,\hat{\bf z}$ pointing downward in the vertical $z$-direction. The domain is periodic in the horizontal directions and has rigid top and bottom walls. Each of the two fluids has an initially uniform density $\rho$ separated by an interface in the middle of the domain (Fig.1 in \citeA{bian2020revisiting}). Prior to instability growth and fluid motion, pressure satisfies hydrostatic equilibrium, $\partial_z p = -\rho g$. Past studies \cite{zhao2018inviscid,zhao2022scale,zhao2025multi} using coarse-graining to analyze scale interactions in these flows had to make a choice regarding the extension of density and pressure beyond the top and bottom walls. The density field was kept constant (zero normal gradient) and pressure was extended linearly, $p=\rho\,g\,z$, in accordance with the hydrostatic relation $\partial_z p=-\rho \,g$ (non-homogeneous Neumann BC). It is straightforward to verify that these choices for $\rho$ and $p$ are the harmonic solutions of the Laplace boundary value problem with the respective BCs for density and pressure.

\subsection{Numerical Implementation}
In practice, the Laplace boundary value problem for the fill domain $\cD_\fil$
is solved on the same grid and stencil as the underlying model, such as that of a general circulation model. Generally, $\nabla^2 u_* = 0$ can be discretized over  $\cD_\fil$ using a standard second-order finite-volume (or finite-difference) scheme, enforcing the boundary conditions weakly by prescribing $u_*$ on the interface $\Bdr$. This yields a sparse, symmetric positive-definite linear system
\be
A u_* = b,
\ee
where $A$ is the discrete Laplace operator restricted to $\cD_\fil$ and $b$
contains the contributions from the imposed boundary values. The system can be
solved with an iterative method such as Gauss-Seidel or conjugate gradients \cite{press2007numerical}, and
because $A$ depends only on the fixed geometry and mask, its structure (or
preconditioner) can be reused across time steps and for multiple fields. Therefore, the harmonic extension adds only a modest numerical cost compared to an existing numerical model.

Below, we choose a standard diffusive relaxation solver \cite{press2007numerical} for solving problem~\eqref{eq:HarmonicExtension}. Instead of solving the steady Laplace problem $\nabla^2 u_* = 0$ in $\cD_\fil$ directly, diffusive relaxation introduces a pseudo-time $\tau$ and considers the artificial diffusion equation 
\be
\frac{\partial u_*}{\partial \tau} = \kappa\,\nabla^2 u_*
\quad \text{in } \cD_\fil,
\lb{eq:diffusion} \ee
with the boundary values on $\Bdr$ held fixed for all $\tau$.  As $\tau\to\infty$, the solution relaxes to a steady state that satisfies the desired Laplace boundary value problem. Here, $\kappa$ is a diffusivity parameter whose value influences only the solver's stability and convergence speed, but not the harmonic solution $u_*$ itself.
Numerically, \eqref{eq:diffusion} amounts to iterating a simple relaxation scheme that mimics diffusion. For example, with an explicit time discretization,
\be
u^{(n+1)}_* = u^{(n)}_* + \Delta\tau\,\kappa\,\nabla^2_h u^{(n)},
\lb{eq:discrete-solver}\ee
where $\nabla^2_h$ is the discrete Laplacian on $\cD_\fil$ and $u^{(n)}$ denotes
the $n$-th pseudo-time iterate. Below, we use a centered $4$-th order finite-difference scheme for $\nabla^2_h$ over land [see eq.~\eqref{eq:4th_order_stencils}]. Ocean grid points are held fixed to their original values at every iteration. The size of the (pseudo-) time-step is constrained by a CFL condition to ensure numerical stability. The harmonic solution $u_*$ is then obtained when the relaxation converges to a steady-state within a prescribed numerical tolerance. Details of the discretization, time-stepping, and convergence criteria are provided in \ref{AppA} and \ref{AppB}.

\section{Application: Land-filling Sea-Surface Temperature}

\subsection{Data}
We use 1-day averaged Level-4 global sea-surface temperature (SST) field from the EU Copernicus Marine Service OSTIA product \cite{Copernicus2020OSTIA}, which provides gap-free analyses at $0.05^\circ$ resolution on a regular latitude-longitude grid by merging satellite and in situ observations through optimal interpolation. The dataset spans the global ocean and is updated in near real time. Here, we use data from May~1,~2025.

\subsection{Initialization Strategy}
\label{initialize}
To speed up convergence of the Laplace solver \eqref{eq:discrete-solver} on the high-resolution $0.05^\circ$ grid, we use a two-level multigrid method by first solving the system on a coarser $0.25^\circ$ grid. 
The coarse grid is initialized using zonal means, which is computationally inexpensive. 
%
For datasets involving several snapshots in time (e.g., a time-series of global SST), the solution from the previous snapshot is used as the initial state, further accelerating convergence while preserving large scale coherence. Implementation details and convergence tests are given in Appendix~B.

\subsection{Boundary Conditions}
\label{BoundaryConditions}

We solve the Laplace boundary value problem~\eqref{eq:HarmonicExtension} over the entire land domain to fill land globally. Because the harmonic extension is determined by boundary conditions, we implement and compare two formulations. Implementation details are provided in \ref{AppA}.

\paragraph*{Dirichlet boundary condition.}
Here, we use ocean cells adjacent to land as a strong (explicit) Dirichlet boundary along the coastline, and no separate treatment or interpolation of interface values is required. This formulation is computationally inexpensive.

\paragraph*{Neumann boundary condition.}
Alternatively, we impose a zero-flux (insulating) boundary condition on the temperature $T$,
\begin{equation}\label{neumann}
\left.\frac{\partial T}{\partial n}\right|_{\partial \cD_{\mathrm{fill}}} = 0,
\end{equation}
at the land-ocean interface while solving the same Laplace equation over interior (non-coastal) land cells. The Neumann boundary condition can be approximated using finite difference schemes of first- or higher-order accuracy. 
In a simple first-order approach, the coastal land temperature is set to the average of the adjacent ocean temperature. 

\subsection{Results}
\begin{figure}
    \centering
    \includegraphics[width=0.99\linewidth]{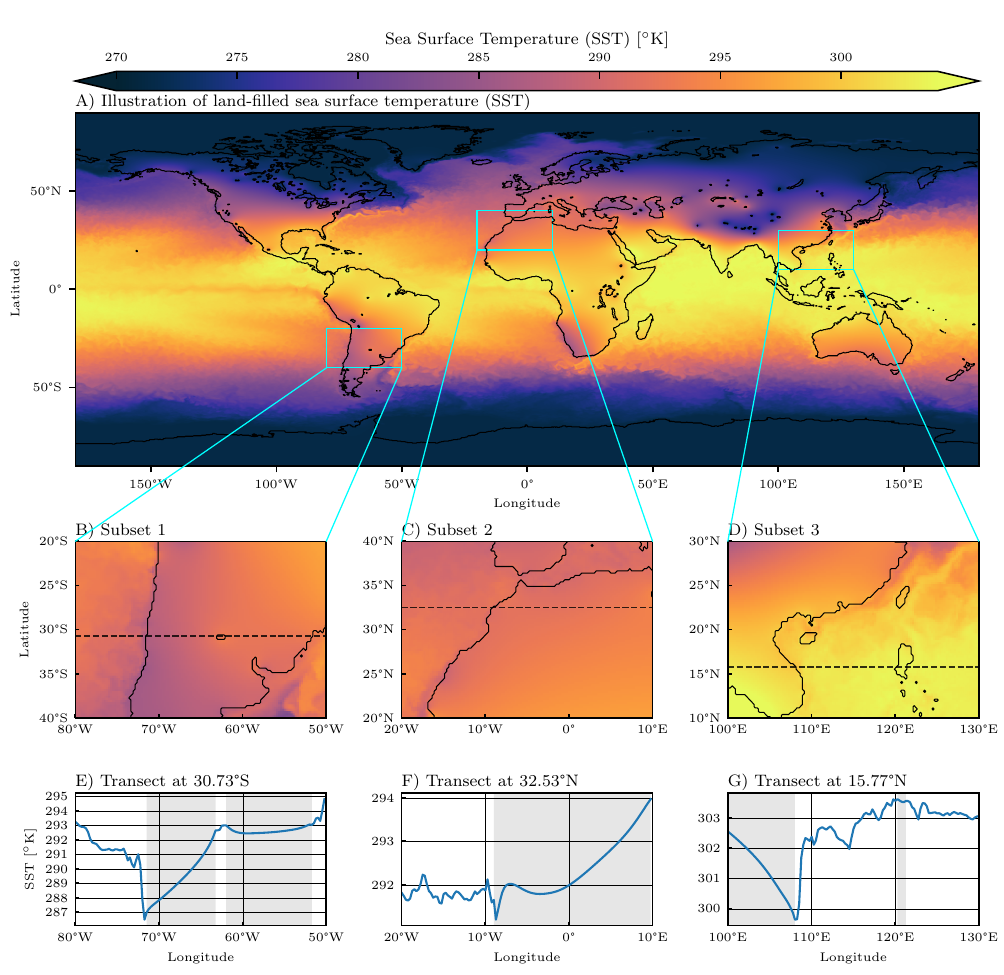}
    \caption{\textbf{Harmonic extension of sea-surface temperature over land.}
    Land values are computed here by enforcing a no-flux boundary condition (\(\partial T/\partial n = 0\)) along coastlines, yielding a smooth, physically consistent harmonic extension of the oceanic SST field across land.
    \emph{Top [A]:} global SST land-filling using a Neumann (no-flux) boundary condition.
    \emph{Middle [B,C,D]:} three representative zooms highlighting coastal smoothness and fidelity.
    \emph{Bottom [E,F,G]:} selected cross-coast transects through filled regions, illustrating zero coast-normal gradients and continuity between ocean and land values.}

    \label{fig:Neumann_zoom}
\end{figure}

Figure~\ref{fig:Neumann_zoom}A illustrates the result of land filling, with convergence metrics presented and discussed in \ref{AppB}. 
The middle row (panels~\ref{fig:Neumann_zoom}B-D) zooms onto coastal regions to highlight smoothness of the land-filling near shore. The dashed lines in Fig.~\ref{fig:Neumann_zoom}B-D indicate the selected transects that are shown (with land shaded) in the bottom row (Fig.~\ref{fig:Neumann_zoom}E-F), highlighting how the harmonic solution fills land smoothly from the adjacent ocean.

\begin{figure}
    \centering
    \includegraphics[width=\linewidth]{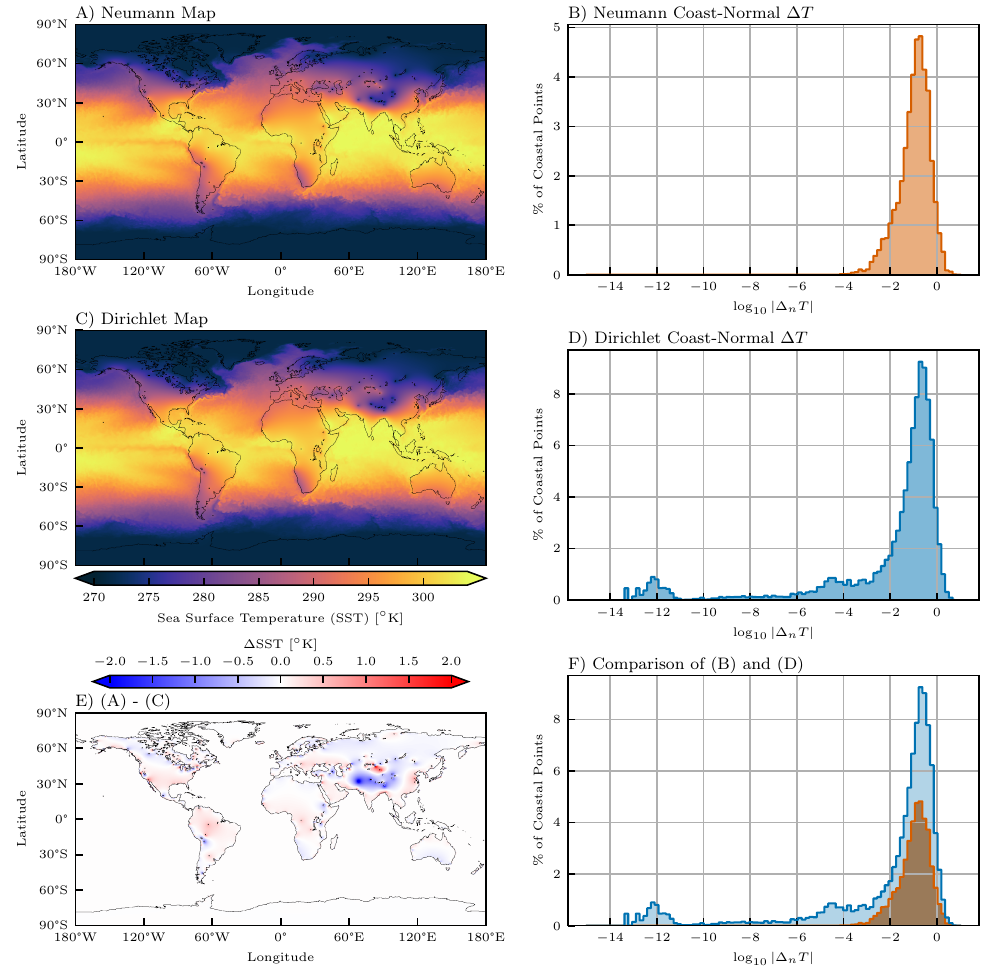}
    \caption{
    \textbf{Comparison of boundary conditions.}
    Harmonically-extended sea surface temperature using [A] Neumann BCs and [C] Dirichlet BCs. 
    (B,D) Histogram of coastal-normal temperature differences for the solution in (A,C), respectively. 
    (E) Difference between the Neumann and Dirichlet harmonic extensions (A-C). 
    (F) Overlay of the two histograms in (B) and (D) for direct comparison.
    Note that the plotted Neumann distribution (orange) does not integrate to unity because of the large number of zero values, which are at $-\infty$ on the log-axis. 
    }
    \label{fig:filledland}
\end{figure}

\paragraph*{Comparing Neumann and Dirichlet boundary conditions.}
Figure~\ref{fig:filledland} compares the results of the Dirichlet and Neumann BCs. Panels~\ref{fig:filledland}A,C show that the Neumann and Dirichlet fields are visually indistinguishable at basin scales, while panel~\ref{fig:filledland}E shows that their differences are small and localized near coastlines and inland water bodies (e.g., lakes).
The right column (panels~\ref{fig:filledland}B,D,F) quantifies the boundary differences via a histogram of coastal-normal temperature differences. 
In the Neumann case (orange curves), the coast-normal differences are not identically zero because of coastal points with multiple land neighbors, but there are substantially fewer points with non-zero coastal-normal gradients using the Neumann BC. 
Note also that the Neumann distribution (orange) does not visually integrate to unity because of the large number of zero values that are not represented on a log-axis. 

\begin{figure}
    \centering
    \includegraphics[width=0.99\linewidth]{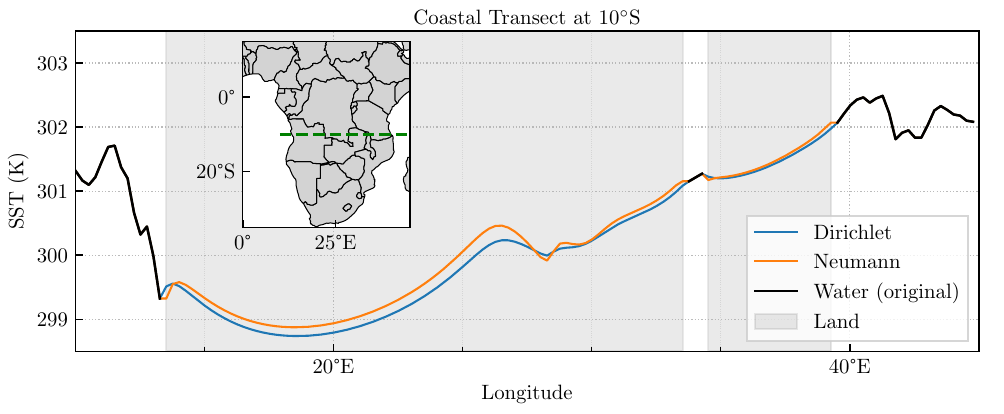}
    \caption{\textbf{Transect comparing Neumann and Dirichlet results}
    The transect follows the 10$^\circ$S parallel from the southeastern Atlantic (off Angola) across southern Africa to the southwestern Indian Ocean (off Tanzania).
    Original (ocean) SST (black lines) together with harmonic extensions using Dirichlet (blue) and Neumann (orange) boundary conditions. 
    Shaded vertical bands mark longitudes corresponding to land. 
    Note that the local maxima at \(\sim26.5^\circ\mathrm{E}\) and \(\sim28.5^\circ\mathrm{E}\) are the imprints of Lake Upemba and Lake Mweru, respectively. 
    The inset map indicates the transect with the dashed green line.
    }
    \label{fig:1D_sst}
\end{figure}

To further compare the two boundary treatments, Figure~\ref{fig:1D_sst} shows a transect at $10^\circ$S extending from the western coast of Angola (south-east Atlantic) to the Indian Ocean (eastern coast of Tanzania). 
The in-set map shows the geographic location of this transect. 
The Dirichlet and Neumann solutions are generally very similar, with differences localized to coasts and in-land bodies of water. 
The transects highlight how those differences can extend in-land through the harmonic extension. 
The pronounced peaks near $26.5^\circ$E and $28.5^\circ$E reflect the influence of Lake Upemba and Lake Mweru, respectively.

\section{Concluding Remarks}
We have presented a harmonic extension framework for extending scalar values to regions beyond the domain boundaries by solving a Laplace boundary value problem consistent with the governing boundary conditions. 
The resulting extension is mathematically well-posed and minimizes spatial variability among all admissible fields that satisfy the prescribed boundary conditions. 
In practice, it yields smooth, physically sensible fields that preserve coastal extrema, is robust to irregular (e.g., polygonal) boundaries, and is straightforward to implement with standard finite-volume or finite-difference solvers and simple relaxation schemes. 
The method is broadly applicable beyond the SST example considered here, including to scale-analysis using Fourier methods or convolution-based coarse-graining, and to training machine-learned parameterizations near complex boundaries.

\paragraph*{An Immersed Boundary Method.}
Although our formulation here has been motivated by diagnostic needs, it also suggests interesting prognostic uses. The harmonic extension naturally embeds the original domain within a larger computational region, with the fill region playing a role analogous to the ``solid'' or ``immersed'' region in immersed boundary methods \cite<IBM; for reviews, see>{peskin2002immersed,mittal2005immersed,griffith2020immersed}. Unlike classic ghost-cell fills, which are typically local algebraic extrapolations, the harmonic extension provides well-posed ghost (land) values that are globally consistent with the physical boundary conditions. The tradeoff is a global elliptic solve each step, but the cost is modest with multigrid methods \cite{brandt1977multi,trottenberg2001multigrid}. In principle, such an embedding could be coupled dynamically to the governing equations or to data-driven closure schemes, enabling treatment of evolving interfaces (e.g., sea-ice or multiphase flows) in a unified framework. A possible usage as an IBM is:
\begin{enumerate}
    \item Define a land-ocean mask.
    \item At each time step, perform a harmonic fill of ghost (land) cells as described here.
    \item Compute right-hand sides, including spatial derivatives, with the standard finite-difference/finite-volume stencils everywhere. Near coasts, stencils that reach into land read the harmonic ghosts.
    \item Time-advance the PDE for the dynamics only over ocean cells.
    \item Refresh land cells via the next harmonic solve.
\end{enumerate}

The cost is one elliptic solve per field per step, which is modest with multigrid methods. Spatial derivatives are evaluated consistently with the same stencils throughout the domain, including near boundaries via ghost-cell values obtained from a principled elliptic extension rather than ad hoc local extrapolations.  Thus the harmonic extension supplies ghost values that are globally optimized (minimum-Dirichlet-energy) and yield physically consistent boundary behavior without body-fitted meshing. This is especially useful for high-order schemes with wide stencils, avoiding special lower-accuracy boundary stencils. From this perspective, spectral methods are essentially highest-order schemes whose effective stencil spans the entire embedding grid. Consequently, our extension framework here opens the possibility of using spherical-harmonic dynamical cores in ocean modeling, with harmonic ghost-cell values over land enforcing coastal boundary conditions and smoothly matching the ocean fields. Moreover, having well-posed ghost-cell values throughout the land mask facilitates multigrid (or multiresolution) flow simulations on complex geometries, where the flow is advanced on a grid hierarchy and coarse levels require ghost information deeper inland to maintain accurate stencils and intergrid (coarse-to-fine) transfers  \cite{trottenberg2001multigrid}. Exploring these prognostic applications, including their stability and accuracy, in full simulations is a promising direction for future work.

\appendix



\section{}\label{AppA}
\setcounter{figure}{0}
\setcounter{equation}{0}
\renewcommand{\thefigure}{A\arabic{figure}}
\renewcommand{\theequation}{A.\arabic{equation}}

\subsection*{Implementation Details: Harmonic Extensions using Dirichlet and Neumann Boundary Conditions}

\begin{enumerate}

\item{\textit{Coarse-to-fine Dirichlet initialization (two-level multigrid diffusion).}
Raw $0.05^\circ$ SST fields defined over the ocean cells are block-averaged onto a coarser grid (here, $0.25^\circ$). 
\begin{itemize}[leftmargin=3cm]
    \item[\textbf{Coarse-level}]{
        On the $0.25^\circ$ grid, we solve $\nabla^2 T = 0$ over land with $T|_{\cD}$ (ocean cells) held fixed by explicitly advancing the diffusion equation $\partial_t T = \nabla^2 T$ until it reaches a steady-state within a prescribed tolerance. 
        Zonal means are used as the initial conditions for simplicity.
        The resulting harmonic field is linearly interpolated to the native $0.05^\circ$ grid.
    }
    \item[\textbf{Fine-level}]{
        Starting from the up-sampled field, we integrate the same diffusion equation again on the $0.05^\circ$ grid with the ocean fixed until convergence.
    }
    \item[Note:]{
        All figures in this paper use $0.25^\circ$ fields to focus on method development, but the full pipeline operates at $0.05^\circ$ as described.
    }
\end{itemize}
}

\item{\textit{Discrete Laplacian operator.}
    Spatial derivatives are computed on a regular latitude-longitude grid using 4th order centered finite difference stencils. 
    The discrete scalar Laplacian on the sphere is expressed as
    \begin{equation}
    \nabla^2 T = 
    \frac{1}{R_E^2}\!\left(
    \frac{1}{\cos^2\phi}\frac{\partial^2 T}{\partial \lambda^2}
    + \frac{\partial^2 T}{\partial \phi^2}
    - \tan\phi\,\frac{\partial T}{\partial \phi}
    \right),
    \end{equation}
    where $\lambda$ and $\phi$ denote longitude and latitude and $R_E$ is Earth’s radius.
    All derivatives use fourth-order centered stencils with periodicity in longitude and mirrored boundaries at the poles, effectively enforcing a no-flux condition there. A small lower bound $\epsilon$ is applied to $\cos\phi$ near the poles for numerical stability.
    The numerical stencils for fourth-order second derivative (\(C^{(2)}_{4th}\)) and first derivative (\(C^{(1)}_{4th}\)) are:
    \begin{equation}
    C^{(2)}_{4th}=\frac{[-1,\,16,\,-30,\,16,\,-1]}{12h^2} \quad \text{and}\quad
    C^{(1)}_{4nd}=\frac{[1,\,-8,\,0,\,8,\,-1]}{12h},
    \label{eq:4th_order_stencils}
    \end{equation}
    where \(h\) denotes the grid spacing.\\[6pt]
}

\item{
    \textit{Dirichlet boundary condition.}
    Dirichlet conditions are imposed by holding the water values constant and evolving the diffusion equation over all land points.\\[6pt]
}

\item{
    \textit{Neumann boundary condition.}

    For each coastal land cell, the normal derivative $\partial T / \partial n$ is approximated using a first-order finite difference from the adjacent ocean neighbor, such that the value on the coastal land cell is set to that of its water neighbor. 
    When a land cell is exposed to more than one ocean neighbor (e.g., at corners or tips), the corresponding values are averaged to obtain a single representative value. 
    This low-order enforcement of zero flux ($\partial T/\partial n \approx 0$) is computationally inexpensive, in contrast to higher order methods that require solving least squares systems [c.f.~\ref{AppC}].\\[6pt]
}

\item{
    \textit{Laplace solution (for land cells).}
    Ocean cells (and coastal land for Neumann BCs) are held fixed in the diffusion solver. 
    The remaining interior land cells are evolved by integrating the diffusion equation:
    \begin{equation}
    \frac{\partial T}{\partial t} = \nabla^2 T,
    \end{equation}
    explicitly in time until the solution reaches steady-state. 
}

\item{\textit{Time stepping and stability.}
    We use explicit Euler integration with a local CFL constraint $\Delta t \le h_{\min}^2/(4\kappa)$, with $\kappa=1$.
    It is worth noting that only the land points are updated (except coasts in the case of Neumann BCs): ocean points remain fixed. 
    In the results presented here, the CFL is fixed at \(0.1\).
    The grid spacing \(h_{\min}=\min(\Delta x,\Delta y)\) is computed in metres using the local grid spacing.
    \begin{equation}
    \Delta x = R_E\cos\phi\,\Delta\lambda_{\mathrm{rad}},\qquad
    \Delta y = R_E\,\Delta\phi_{\mathrm{rad}}.
    \end{equation}
}
\end{enumerate}

\section{}\label{AppB}
\setcounter{figure}{0}
\setcounter{equation}{0}
\renewcommand{\thefigure}{B\arabic{figure}}
\renewcommand{\theequation}{B.\arabic{equation}}

\paragraph*{Convergence Metrics}

\begin{figure}
    \centering
    \includegraphics[scale=1]{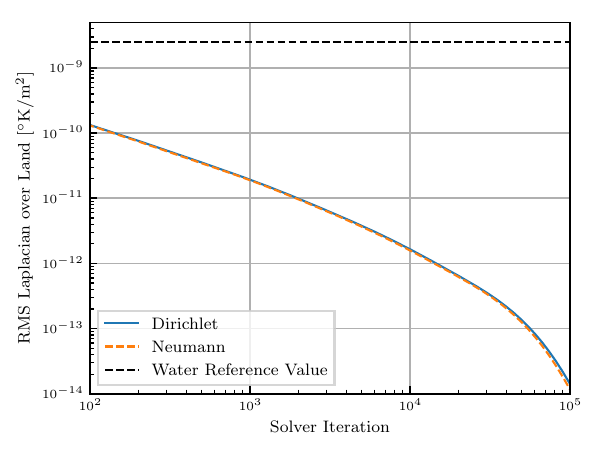}
    \caption{
        \textbf{Convergence of the RMS Laplacian} 
        For both the Dirichlet (blue) and Neumann (orange) boundary conditions, we compute the RMS Laplacian of the land-filled scalar throughout the iterative solver. Convergence is reached once the RMS value falls below a desired tolerance. 
        For reference, the dashed black line in Fig.~\ref{fig:app:convergence_1} shows the RMS Laplacian evaluated over water.
    }
    \label{fig:app:convergence_1}
\end{figure}

To track the convergence of the iterative solver, we consider the root-mean-squared (RMS) Laplacian of the scalar over land cells, \(\cD_{\mathrm{land}}\). For Dirichlet boundary conditions, \(\cD_{\mathrm{land}}\) includes all land cells, while under Neumann conditions \(\cD_{\mathrm{land}}\) excludes coastal cells.
\begin{equation}
    \mathrm{RMS}(\nabla^2 s) :=\frac{1}{\absv{\cD_{\mathrm{land}}}}\int_{\cD_{\mathrm{land}}}\nabla^2s\,\mathrm{d}A
\end{equation}
Figure~\ref{fig:app:convergence_1} presents the RMS Laplacian for both the Dirichlet and Neumann cases. 
Both cases demonstrate similar convergence rates, and show a decrease in the RMS Laplacian by four orders of magnitude from initalization. 
For reference, the dashed black line in Fig.~\ref{fig:app:convergence_1} shows the RMS Laplacian evaluated over water.

\section{}\label{AppC}
\setcounter{figure}{0}
\setcounter{equation}{0}
\renewcommand{\thefigure}{C\arabic{figure}}
\renewcommand{\theequation}{C.\arabic{equation}}

\subsection*{Higher Order Neumann Boundary Condition}


Enforcing the Neumann boundary condition with a higher order derivative introduces additional computational complexity. 
Here we outline the process for enforcing the second-order centered derivative to be zero on coastal points.
The corresponding normal gradient constraints can be solved using a minimum change projection (large sparse least-squares system, LSMR; see \citeA{fong2011lsmr}) which modifies coastal land values to drive the normal gradients toward zero, while keeping ocean points unchanged.
\begin{enumerate}
            
            
\item{
    \textit{Discrete derivative operators (4th-order interior, 2nd-order near boundaries) and spherical Laplacian.}
    All spatial second derivatives are computed directly using second-derivative finite-difference stencils on a latitude-longitude grid, without composing first-derivative operators.
    We employ 4th order centered stencils (eq.~\eqref{eq:4th_order_stencils}) in the interior and 2nd-order centered stencils (\(C^{(1)}_{2nd}\), below) near boundaries to impose Neumann BCs.
    \begin{equation}
    C^{(1)}_{2nd}=\frac{[-1,\,0,\,1]}{h}
    \end{equation}
    Sparse derivative matrices $C_{\lambda}^{(2)}$ and $C_{\phi}^{(2)}$ are assembled for longitude and latitude, respectively, which act on inputs of size $N_{\mathrm{lon}}$ and $ N_{\mathrm{lat}}$. 
    Two-dimensional operators acting on the full field of size $N_{\mathrm{lat}}\times N_{\mathrm{lon}}$ are built by Kronecker products:
    \begin{equation}
    C_{\lambda}^{(2,Kr)} = I_{\phi}\!\otimes C_{\lambda}^{(2)}, 
    \qquad
    C_{\phi}^{(2,Kr)} = C_{\phi}^{(2)}\!\otimes I_{\lambda},
    \end{equation}
    where \(I_\lambda\) and \(I_\phi\) denote the identity matrices of size $(N_{\mathrm{lon}},N_{\mathrm{lon}})$ and $(N_{\mathrm{lat}},N_{\mathrm{lat}})$, respectively.
    The spherical scalar Laplacian is expressed as:
    
    \begin{equation}
        \nabla^2 T =
    \frac{1}{R_E^2}
    \paren{
    \frac{1}{\cos^2\phi}\frac{\partial^2 T}{\partial \lambda^2}
    + \frac{\partial^2 T}{\partial \phi^2}
    - \tan\phi\,\frac{\partial T}{\partial \phi}
    }.
    \end{equation}
    In discrete matrix form, this becomes
    \begin{equation}
    \nabla^2 T \approx
    \frac{1}{R_E^2}
    \paren{
    D_{\cos}^{-2} C_{\lambda}^{(2,Kr)} T
    + C_{\phi}^{(2,Kr)} T
    - D_{\tan\phi}\,C_{\phi}^{(1,Kr)} T
    },
    \end{equation}
    where $D_{\cos}=\mathrm{diag}(\cos\phi)$ and 
    $D_{\tan\phi}=\mathrm{diag}(\tan\phi)$.
    }
    
\item{ \textit{Coastline masks for boundary conditions.}
    Boolean masks $L_x$ and $L_y$ mark land cells adjacent to ocean in the east-west and north-south directions, respectively. 
    To support centered stencils while only modifying land, we impose the boundary conditions within a two-cell dilation of these coastal sets.
}

\item{ \textit{Neumann enforcement via projected least squares (centered 2nd order).}
    We impose the no-flux condition by minimizing the coast-normal components of the gradient using centered second-order first-derivative operators:
    \begin{equation}
    D_{\mathrm{neu}} \;=\;
    \begin{bmatrix}
    \mathcal{R}_x\, C_{\lambda}^{(1,2D)} \\
    \mathcal{R}_y\, C_{\phi}^{(1,2D)}
    \end{bmatrix}, 
    \qquad r = D_{\mathrm{neu}}\,\mathrm{vec}(T),
    \end{equation}
    where $\mathcal{R}_x,\mathcal{R}_y$ restrict rows to the (dilated) sets in $L_x,L_y$. 
    At each relaxation step we solve a damped least-squares problem (LSMR) for a minimal correction \(\Delta\) supported only on the dilated coastal land band:
    \begin{equation}
    \min_{\Delta_{\mathrm{land}}}\ 
    \absv{ D_{\mathrm{neu}}\big(T - \Delta\big) }_2^2
    \;+\; \mu^2 \absv{\Delta_{\mathrm{land}}}_2^2,
    \quad \text{s.t. }\Delta_{\mathrm{ocean}}=0,
    \end{equation}
    with small Tikhonov parameter $\mu$, loose tolerances (e.g., $10^{-4}$), and a modest iteration cap (e.g., 40). 
    Warm starts use the previous $\Delta$ to improve convergence. 
    The update is $T - \Delta \rightarrow T$. All other computational steps follow the same procedure described in Appendix~A.

}
\end{enumerate}

\section*{Open Research Section}
The SST data is available publicly from the EU Copernicus Marine Service \cite{Copernicus2020OSTIA}. 
All analysis and figure generation Jupyter notebooks will be made publicly available through a Zenodo archive upon acceptance of the manuscript.

\acknowledgments
This research was supported by NSF OCE-2123496, OCE-2446475. HA was also supported by US NSF grants TI-2449340, PHY-2020249, PHY-2206380, US DOE grants DE-SC0020229, and by 
US NNSA grants DE-NA0004144, 
and DE-NA0004134.
\bibliography{refs}

\begin{thebibliography}{}

\bibitem [\protect \citeauthoryear {%
Aluie%
}{%
Aluie%
}{%
{\protect \APACyear {2017}}%
}]{%
aluie2017coarse}
\APACinsertmetastar {%
aluie2017coarse}%
\begin{APACrefauthors}%
Aluie, H.%
\end{APACrefauthors}%
\unskip\
\newblock
\APACrefYearMonthDay{2017}{}{}.
\newblock
{\BBOQ}\APACrefatitle {Coarse-grained incompressible magnetohydrodynamics:
  analyzing the turbulent cascades} {Coarse-grained incompressible
  magnetohydrodynamics: analyzing the turbulent cascades}.{\BBCQ}
\newblock
\APACjournalVolNumPages{New Journal of Physics}{19}{2}{025008}.
\PrintBackRefs{\CurrentBib}

\bibitem [\protect \citeauthoryear {%
Aluie%
}{%
Aluie%
}{%
{\protect \APACyear {2019}}%
}]{%
aluie2019conv}
\APACinsertmetastar {%
aluie2019conv}%
\begin{APACrefauthors}%
Aluie, H.%
\end{APACrefauthors}%
\unskip\
\newblock
\APACrefYearMonthDay{2019}{}{}.
\newblock
{\BBOQ}\APACrefatitle {Convolutions on the sphere: Commutation with
  differential operators} {Convolutions on the sphere: Commutation with
  differential operators}.{\BBCQ}
\newblock
\APACjournalVolNumPages{GEM-International Journal on Geomathematics}{10}{1}{9}.
\PrintBackRefs{\CurrentBib}

\bibitem [\protect \citeauthoryear {%
Aluie%
, Hecht%
\BCBL {}\ \BBA {} Vallis%
}{%
Aluie%
\ \protect \BOthers {.}}{%
{\protect \APACyear {2018}}%
}]{%
aluie2018mapping}
\APACinsertmetastar {%
aluie2018mapping}%
\begin{APACrefauthors}%
Aluie, H.%
, Hecht, M.%
\BCBL {}\ \BBA {} Vallis, G\BPBI K.%
\end{APACrefauthors}%
\unskip\
\newblock
\APACrefYearMonthDay{2018}{}{}.
\newblock
{\BBOQ}\APACrefatitle {Mapping the energy cascade in the North Atlantic Ocean:
  The coarse-graining approach} {Mapping the energy cascade in the north
  atlantic ocean: The coarse-graining approach}.{\BBCQ}
\newblock
\APACjournalVolNumPages{Journal of Physical Oceanography}{48}{2}{225--244}.
\PrintBackRefs{\CurrentBib}

\bibitem [\protect \citeauthoryear {%
Axler%
, Bourdon%
\BCBL {}\ \BBA {} Wade%
}{%
Axler%
\ \protect \BOthers {.}}{%
{\protect \APACyear {2013}}%
}]{%
axler2013harmonic}
\APACinsertmetastar {%
axler2013harmonic}%
\begin{APACrefauthors}%
Axler, S.%
, Bourdon, P.%
\BCBL {}\ \BBA {} Wade, R.%
\end{APACrefauthors}%
\unskip\
\newblock
\APACrefYear{2013}.
\newblock
\APACrefbtitle {Harmonic function theory} {Harmonic function theory}\
  (\BVOL~137).
\newblock
\APACaddressPublisher{}{Springer Science \& Business Media}.
\PrintBackRefs{\CurrentBib}

\bibitem [\protect \citeauthoryear {%
Barkan%
, Srinivasan%
\BCBL {}\ \BBA {} McWilliams%
}{%
Barkan%
\ \protect \BOthers {.}}{%
{\protect \APACyear {2024}}%
}]{%
barkan2024eddy}
\APACinsertmetastar {%
barkan2024eddy}%
\begin{APACrefauthors}%
Barkan, R.%
, Srinivasan, K.%
\BCBL {}\ \BBA {} McWilliams, J\BPBI C.%
\end{APACrefauthors}%
\unskip\
\newblock
\APACrefYearMonthDay{2024}{}{}.
\newblock
{\BBOQ}\APACrefatitle {Eddy--internal wave interactions: Stimulated cascades in
  cross-scale kinetic energy and enstrophy fluxes} {Eddy--internal wave
  interactions: Stimulated cascades in cross-scale kinetic energy and enstrophy
  fluxes}.{\BBCQ}
\newblock
\APACjournalVolNumPages{Journal of Physical Oceanography}{54}{6}{1309--1326}.
\PrintBackRefs{\CurrentBib}

\bibitem [\protect \citeauthoryear {%
Berger%
\ \BBA {} Field%
}{%
Berger%
\ \BBA {} Field%
}{%
{\protect \APACyear {1984}}%
}]{%
berger1984topological}
\APACinsertmetastar {%
berger1984topological}%
\begin{APACrefauthors}%
Berger, M\BPBI A.%
\BCBT {}\ \BBA {} Field, G\BPBI B.%
\end{APACrefauthors}%
\unskip\
\newblock
\APACrefYearMonthDay{1984}{}{}.
\newblock
{\BBOQ}\APACrefatitle {The topological properties of magnetic helicity} {The
  topological properties of magnetic helicity}.{\BBCQ}
\newblock
\APACjournalVolNumPages{Journal of Fluid Mechanics}{147}{}{133--148}.
\PrintBackRefs{\CurrentBib}

\bibitem [\protect \citeauthoryear {%
Bian%
, Aluie%
, Zhao%
, Zhang%
\BCBL {}\ \BBA {} Livescu%
}{%
Bian%
\ \protect \BOthers {.}}{%
{\protect \APACyear {2020}}%
}]{%
bian2020revisiting}
\APACinsertmetastar {%
bian2020revisiting}%
\begin{APACrefauthors}%
Bian, X.%
, Aluie, H.%
, Zhao, D.%
, Zhang, H.%
\BCBL {}\ \BBA {} Livescu, D.%
\end{APACrefauthors}%
\unskip\
\newblock
\APACrefYearMonthDay{2020}{}{}.
\newblock
{\BBOQ}\APACrefatitle {Revisiting the late-time growth of single-mode
  Rayleigh--Taylor instability and the role of vorticity} {Revisiting the
  late-time growth of single-mode rayleigh--taylor instability and the role of
  vorticity}.{\BBCQ}
\newblock
\APACjournalVolNumPages{Physica D: Nonlinear Phenomena}{403}{}{132250}.
\PrintBackRefs{\CurrentBib}

\bibitem [\protect \citeauthoryear {%
Bose%
\ \BBA {} Park%
}{%
Bose%
\ \BBA {} Park%
}{%
{\protect \APACyear {2018}}%
}]{%
bose2018wall}
\APACinsertmetastar {%
bose2018wall}%
\begin{APACrefauthors}%
Bose, S\BPBI T.%
\BCBT {}\ \BBA {} Park, G\BPBI I.%
\end{APACrefauthors}%
\unskip\
\newblock
\APACrefYearMonthDay{2018}{}{}.
\newblock
{\BBOQ}\APACrefatitle {Wall-modeled large-eddy simulation for complex turbulent
  flows} {Wall-modeled large-eddy simulation for complex turbulent
  flows}.{\BBCQ}
\newblock
\APACjournalVolNumPages{Annual review of fluid mechanics}{50}{}{535--561}.
\PrintBackRefs{\CurrentBib}

\bibitem [\protect \citeauthoryear {%
Boyd%
}{%
Boyd%
}{%
{\protect \APACyear {2001}}%
}]{%
boyd2001chebyshev}
\APACinsertmetastar {%
boyd2001chebyshev}%
\begin{APACrefauthors}%
Boyd, J\BPBI P.%
\end{APACrefauthors}%
\unskip\
\newblock
\APACrefYear{2001}.
\newblock
\APACrefbtitle {Chebyshev and Fourier spectral methods} {Chebyshev and fourier
  spectral methods}.
\newblock
\APACaddressPublisher{}{Courier Corporation}.
\PrintBackRefs{\CurrentBib}

\bibitem [\protect \citeauthoryear {%
Boyd%
}{%
Boyd%
}{%
{\protect \APACyear {2002}}%
}]{%
boyd2002comparison}
\APACinsertmetastar {%
boyd2002comparison}%
\begin{APACrefauthors}%
Boyd, J\BPBI P.%
\end{APACrefauthors}%
\unskip\
\newblock
\APACrefYearMonthDay{2002}{}{}.
\newblock
{\BBOQ}\APACrefatitle {A comparison of numerical algorithms for Fourier
  extension of the first, second, and third kinds} {A comparison of numerical
  algorithms for fourier extension of the first, second, and third
  kinds}.{\BBCQ}
\newblock
\APACjournalVolNumPages{Journal of Computational Physics}{178}{1}{118--160}.
\PrintBackRefs{\CurrentBib}

\bibitem [\protect \citeauthoryear {%
Brandt%
}{%
Brandt%
}{%
{\protect \APACyear {1977}}%
}]{%
brandt1977multi}
\APACinsertmetastar {%
brandt1977multi}%
\begin{APACrefauthors}%
Brandt, A.%
\end{APACrefauthors}%
\unskip\
\newblock
\APACrefYearMonthDay{1977}{}{}.
\newblock
{\BBOQ}\APACrefatitle {Multi-level adaptive solutions to boundary-value
  problems} {Multi-level adaptive solutions to boundary-value problems}.{\BBCQ}
\newblock
\APACjournalVolNumPages{Mathematics of computation}{31}{138}{333--390}.
\PrintBackRefs{\CurrentBib}

\bibitem [\protect \citeauthoryear {%
Buzzicotti%
, Storer%
, Griffies%
\BCBL {}\ \BBA {} Aluie%
}{%
Buzzicotti%
\ \protect \BOthers {.}}{%
{\protect \APACyear {2022}}%
}]{%
buzzicotti2022coarse}
\APACinsertmetastar {%
buzzicotti2022coarse}%
\begin{APACrefauthors}%
Buzzicotti, M.%
, Storer, B\BPBI A.%
, Griffies, S\BPBI M.%
\BCBL {}\ \BBA {} Aluie, H.%
\end{APACrefauthors}%
\unskip\
\newblock
\APACrefYearMonthDay{2022}{}{}.
\newblock
{\BBOQ}\APACrefatitle {A coarse-grained decomposition of surface geostrophic
  kinetic energy in the global ocean} {A coarse-grained decomposition of
  surface geostrophic kinetic energy in the global ocean}.{\BBCQ}
\newblock
\APACjournalVolNumPages{Authorea Preprints}{}{}{}.
\PrintBackRefs{\CurrentBib}

\bibitem [\protect \citeauthoryear {%
Buzzicotti%
, Storer%
, Khatri%
, Griffies%
\BCBL {}\ \BBA {} Aluie%
}{%
Buzzicotti%
\ \protect \BOthers {.}}{%
{\protect \APACyear {2023}}%
}]{%
buzzicotti2023spatio}
\APACinsertmetastar {%
buzzicotti2023spatio}%
\begin{APACrefauthors}%
Buzzicotti, M.%
, Storer, B\BPBI A.%
, Khatri, H.%
, Griffies, S\BPBI M.%
\BCBL {}\ \BBA {} Aluie, H.%
\end{APACrefauthors}%
\unskip\
\newblock
\APACrefYearMonthDay{2023}{}{}.
\newblock
{\BBOQ}\APACrefatitle {Spatio-temporal coarse-graining decomposition of the
  global ocean geostrophic kinetic energy} {Spatio-temporal coarse-graining
  decomposition of the global ocean geostrophic kinetic energy}.{\BBCQ}
\newblock
\APACjournalVolNumPages{Journal of Advances in Modeling Earth
  Systems}{15}{6}{e2023MS003693}.
\PrintBackRefs{\CurrentBib}

\bibitem [\protect \citeauthoryear {%
{Copernicus Marine Service}%
}{%
{Copernicus Marine Service}%
}{%
{\protect \APACyear {2020}}%
}]{%
Copernicus2020OSTIA}
\APACinsertmetastar {%
Copernicus2020OSTIA}%
\begin{APACrefauthors}%
{Copernicus Marine Service}.%
\end{APACrefauthors}%
\unskip\
\newblock
\APACrefYearMonthDay{2020}{}{}.
\newblock
\APACrefbtitle {Global Ocean OSTIA Sea Surface Temperature and Sea Ice Analysis
  (SST\_GLO\_SST\_L4\_NRT\_OBSERVATIONS\_010\_001).} {Global ocean ostia sea
  surface temperature and sea ice analysis
  (sst\_glo\_sst\_l4\_nrt\_observations\_010\_001).}
\newblock
\APACaddressPublisher{}{Copernicus Marine Environment Monitoring Service}.
\newblock
\begin{APACrefURL}
  \url{https://data.marine.copernicus.eu/product/SST_GLO_SST_L4_NRT_OBSERVATIONS_010_001/description}
  \end{APACrefURL}
\newblock
\APACrefnote{Accessed: 2025-06-01}
\newblock
\begin{APACrefDOI} \doi{10.48670/moi-00165} \end{APACrefDOI}
\PrintBackRefs{\CurrentBib}

\bibitem [\protect \citeauthoryear {%
Danilov%
, Juricke%
, Nowak%
, Sidorenko%
\BCBL {}\ \BBA {} Wang%
}{%
Danilov%
\ \protect \BOthers {.}}{%
{\protect \APACyear {2024}}%
}]{%
danilov2024extracting}
\APACinsertmetastar {%
danilov2024extracting}%
\begin{APACrefauthors}%
Danilov, S.%
, Juricke, S.%
, Nowak, K.%
, Sidorenko, D.%
\BCBL {}\ \BBA {} Wang, Q.%
\end{APACrefauthors}%
\unskip\
\newblock
\APACrefYearMonthDay{2024}{}{}.
\newblock
{\BBOQ}\APACrefatitle {Extracting spatial spectra using coarse-graining based
  on implicit filters} {Extracting spatial spectra using coarse-graining based
  on implicit filters}.{\BBCQ}
\newblock
\APACjournalVolNumPages{Journal of Advances in Modeling Earth
  Systems}{16}{5}{e2023MS003946}.
\PrintBackRefs{\CurrentBib}

\bibitem [\protect \citeauthoryear {%
Deike%
}{%
Deike%
}{%
{\protect \APACyear {2022}}%
}]{%
deike2022mass}
\APACinsertmetastar {%
deike2022mass}%
\begin{APACrefauthors}%
Deike, L.%
\end{APACrefauthors}%
\unskip\
\newblock
\APACrefYearMonthDay{2022}{}{}.
\newblock
{\BBOQ}\APACrefatitle {Mass transfer at the ocean--atmosphere interface: the
  role of wave breaking, droplets, and bubbles} {Mass transfer at the
  ocean--atmosphere interface: the role of wave breaking, droplets, and
  bubbles}.{\BBCQ}
\newblock
\APACjournalVolNumPages{Annual Review of Fluid Mechanics}{54}{1}{191--224}.
\PrintBackRefs{\CurrentBib}

\bibitem [\protect \citeauthoryear {%
Evans%
}{%
Evans%
}{%
{\protect \APACyear {2022}}%
}]{%
evans2022partial}
\APACinsertmetastar {%
evans2022partial}%
\begin{APACrefauthors}%
Evans, L\BPBI C.%
\end{APACrefauthors}%
\unskip\
\newblock
\APACrefYear{2022}.
\newblock
\APACrefbtitle {Partial differential equations} {Partial differential
  equations}\ (\BVOL~19).
\newblock
\APACaddressPublisher{}{American mathematical society}.
\PrintBackRefs{\CurrentBib}

\bibitem [\protect \citeauthoryear {%
Eyink%
\ \BBA {} Aluie%
}{%
Eyink%
\ \BBA {} Aluie%
}{%
{\protect \APACyear {2009}}%
}]{%
EyinkAluie09}
\APACinsertmetastar {%
EyinkAluie09}%
\begin{APACrefauthors}%
Eyink, G.%
\BCBT {}\ \BBA {} Aluie, H.%
\end{APACrefauthors}%
\unskip\
\newblock
\APACrefYearMonthDay{2009}{{\APACmonth{11}}}{}.
\newblock
{\BBOQ}\APACrefatitle {{Localness of energy cascade in hydrodynamic turbulence.
  I. Smooth coarse graining}} {{Localness of energy cascade in hydrodynamic
  turbulence. I. Smooth coarse graining}}.{\BBCQ}
\newblock
\APACjournalVolNumPages{Phys. Fluids}{21}{11}{115107}.
\PrintBackRefs{\CurrentBib}

\bibitem [\protect \citeauthoryear {%
{Eyink}%
}{%
{Eyink}%
}{%
{\protect \APACyear {1995}}%
}]{%
Eyink95}
\APACinsertmetastar {%
Eyink95}%
\begin{APACrefauthors}%
{Eyink}, G\BPBI L.%
\end{APACrefauthors}%
\unskip\
\newblock
\APACrefYearMonthDay{1995}{}{}.
\newblock
{\BBOQ}\APACrefatitle {{Local energy flux and the refined similarity
  hypothesis}} {{Local energy flux and the refined similarity
  hypothesis}}.{\BBCQ}
\newblock
\APACjournalVolNumPages{J. Stat. Phys.}{78}{}{335-351}.
\newblock
\begin{APACrefDOI} \doi{10.1007/BF02183352} \end{APACrefDOI}
\PrintBackRefs{\CurrentBib}

\bibitem [\protect \citeauthoryear {%
{Finn}%
\ \BBA {} {Antonsen}%
}{%
{Finn}%
\ \BBA {} {Antonsen}%
}{%
{\protect \APACyear {1985}}%
}]{%
FinnAntonsen1985}
\APACinsertmetastar {%
FinnAntonsen1985}%
\begin{APACrefauthors}%
{Finn}, J\BPBI M.%
\BCBT {}\ \BBA {} {Antonsen}, T\BPBI M.%
\end{APACrefauthors}%
\unskip\
\newblock
\APACrefYearMonthDay{1985}{}{}.
\newblock
{\BBOQ}\APACrefatitle {{Magnetic helicity: What is it and what is it good
  for?}} {{Magnetic helicity: What is it and what is it good for?}}{\BBCQ}
\newblock
\APACjournalVolNumPages{Comments Plasma Phys. Contr. Fusion}{9}{}{111-126}.
\newblock
\begin{APACrefDOI} \doi{10.1016/j.physd.2006.08.009} \end{APACrefDOI}
\PrintBackRefs{\CurrentBib}

\bibitem [\protect \citeauthoryear {%
Fong%
\ \BBA {} Saunders%
}{%
Fong%
\ \BBA {} Saunders%
}{%
{\protect \APACyear {2011}}%
}]{%
fong2011lsmr}
\APACinsertmetastar {%
fong2011lsmr}%
\begin{APACrefauthors}%
Fong, D\BPBI C\BHBI L.%
\BCBT {}\ \BBA {} Saunders, M.%
\end{APACrefauthors}%
\unskip\
\newblock
\APACrefYearMonthDay{2011}{}{}.
\newblock
{\BBOQ}\APACrefatitle {LSMR: An iterative algorithm for sparse least-squares
  problems} {Lsmr: An iterative algorithm for sparse least-squares
  problems}.{\BBCQ}
\newblock
\APACjournalVolNumPages{SIAM Journal on Scientific
  Computing}{33}{5}{2950--2971}.
\PrintBackRefs{\CurrentBib}

\bibitem [\protect \citeauthoryear {%
Germano%
}{%
Germano%
}{%
{\protect \APACyear {1992}}%
}]{%
germano1992turbulence}
\APACinsertmetastar {%
germano1992turbulence}%
\begin{APACrefauthors}%
Germano, M.%
\end{APACrefauthors}%
\unskip\
\newblock
\APACrefYearMonthDay{1992}{}{}.
\newblock
{\BBOQ}\APACrefatitle {Turbulence: the filtering approach} {Turbulence: the
  filtering approach}.{\BBCQ}
\newblock
\APACjournalVolNumPages{Journal of Fluid Mechanics}{238}{}{325--336}.
\PrintBackRefs{\CurrentBib}

\bibitem [\protect \citeauthoryear {%
Gilbarg%
, Trudinger%
, Gilbarg%
\BCBL {}\ \BBA {} Trudinger%
}{%
Gilbarg%
\ \protect \BOthers {.}}{%
{\protect \APACyear {1977}}%
}]{%
gilbarg1977elliptic}
\APACinsertmetastar {%
gilbarg1977elliptic}%
\begin{APACrefauthors}%
Gilbarg, D.%
, Trudinger, N\BPBI S.%
, Gilbarg, D.%
\BCBL {}\ \BBA {} Trudinger, N.%
\end{APACrefauthors}%
\unskip\
\newblock
\APACrefYear{1977}.
\newblock
\APACrefbtitle {Elliptic partial differential equations of second order}
  {Elliptic partial differential equations of second order}\ (\BVOL~224)\
  (\BNUM~2).
\newblock
\APACaddressPublisher{}{Springer}.
\PrintBackRefs{\CurrentBib}

\bibitem [\protect \citeauthoryear {%
Griffith%
\ \BBA {} Patankar%
}{%
Griffith%
\ \BBA {} Patankar%
}{%
{\protect \APACyear {2020}}%
}]{%
griffith2020immersed}
\APACinsertmetastar {%
griffith2020immersed}%
\begin{APACrefauthors}%
Griffith, B\BPBI E.%
\BCBT {}\ \BBA {} Patankar, N\BPBI A.%
\end{APACrefauthors}%
\unskip\
\newblock
\APACrefYearMonthDay{2020}{}{}.
\newblock
{\BBOQ}\APACrefatitle {Immersed methods for fluid--structure interaction}
  {Immersed methods for fluid--structure interaction}.{\BBCQ}
\newblock
\APACjournalVolNumPages{Annual review of fluid mechanics}{52}{1}{421--448}.
\PrintBackRefs{\CurrentBib}

\bibitem [\protect \citeauthoryear {%
Grooms%
\ \protect \BOthers {.}}{%
Grooms%
\ \protect \BOthers {.}}{%
{\protect \APACyear {2021}}%
}]{%
grooms2021diffusion}
\APACinsertmetastar {%
grooms2021diffusion}%
\begin{APACrefauthors}%
Grooms, I.%
, Loose, N.%
, Abernathey, R.%
, Steinberg, J.%
, Bachman, S\BPBI D.%
, Marques, G.%
\BDBL {}Yankovsky, E.%
\end{APACrefauthors}%
\unskip\
\newblock
\APACrefYearMonthDay{2021}{}{}.
\newblock
{\BBOQ}\APACrefatitle {Diffusion-based smoothers for spatial filtering of
  gridded geophysical data} {Diffusion-based smoothers for spatial filtering of
  gridded geophysical data}.{\BBCQ}
\newblock
\APACjournalVolNumPages{Journal of Advances in Modeling Earth
  Systems}{13}{9}{e2021MS002552}.
\PrintBackRefs{\CurrentBib}

\bibitem [\protect \citeauthoryear {%
Hadamard%
}{%
Hadamard%
}{%
{\protect \APACyear {1902}}%
}]{%
hadamard1902problemes}
\APACinsertmetastar {%
hadamard1902problemes}%
\begin{APACrefauthors}%
Hadamard, J.%
\end{APACrefauthors}%
\unskip\
\newblock
\APACrefYearMonthDay{1902}{}{}.
\newblock
{\BBOQ}\APACrefatitle {Sur les probl{\`e}mes aux d{\'e}riv{\'e}es partielles et
  leur signification physique} {Sur les probl{\`e}mes aux d{\'e}riv{\'e}es
  partielles et leur signification physique}.{\BBCQ}
\newblock
\APACjournalVolNumPages{Princeton university bulletin}{}{}{49--52}.
\PrintBackRefs{\CurrentBib}

\bibitem [\protect \citeauthoryear {%
Juricke%
, Bellinghausen%
, Danilov%
, Kutsenko%
\BCBL {}\ \BBA {} Oliver%
}{%
Juricke%
\ \protect \BOthers {.}}{%
{\protect \APACyear {2023}}%
}]{%
juricke2023scale}
\APACinsertmetastar {%
juricke2023scale}%
\begin{APACrefauthors}%
Juricke, S.%
, Bellinghausen, K.%
, Danilov, S.%
, Kutsenko, A.%
\BCBL {}\ \BBA {} Oliver, M.%
\end{APACrefauthors}%
\unskip\
\newblock
\APACrefYearMonthDay{2023}{}{}.
\newblock
{\BBOQ}\APACrefatitle {Scale analysis on unstructured grids: Kinetic energy and
  dissipation power spectra on triangular meshes} {Scale analysis on
  unstructured grids: Kinetic energy and dissipation power spectra on
  triangular meshes}.{\BBCQ}
\newblock
\APACjournalVolNumPages{Journal of Advances in Modeling Earth
  Systems}{15}{1}{e2022MS003280}.
\PrintBackRefs{\CurrentBib}

\bibitem [\protect \citeauthoryear {%
Khatri%
\ \protect \BOthers {.}}{%
Khatri%
\ \protect \BOthers {.}}{%
{\protect \APACyear {2024}}%
}]{%
khatri2024scale}
\APACinsertmetastar {%
khatri2024scale}%
\begin{APACrefauthors}%
Khatri, H.%
, Griffies, S\BPBI M.%
, Storer, B\BPBI A.%
, Buzzicotti, M.%
, Aluie, H.%
, Sonnewald, M.%
\BDBL {}Shao, A\BPBI E.%
\end{APACrefauthors}%
\unskip\
\newblock
\APACrefYearMonthDay{2024}{}{}.
\newblock
{\BBOQ}\APACrefatitle {A scale-dependent analysis of the barotropic vorticity
  budget in a global ocean simulation} {A scale-dependent analysis of the
  barotropic vorticity budget in a global ocean simulation}.{\BBCQ}
\newblock
\APACjournalVolNumPages{Journal of Advances in Modeling Earth Systems}{}{}{}.
\newblock
\APACrefnote{under review}
\newblock
\begin{APACrefDOI} \doi{10.22541/essoar.168394747.71837050/v1} \end{APACrefDOI}
\PrintBackRefs{\CurrentBib}

\bibitem [\protect \citeauthoryear {%
Kouhen%
, Storer%
, Aluie%
, Marshall%
\BCBL {}\ \BBA {} Christensen%
}{%
Kouhen%
\ \protect \BOthers {.}}{%
{\protect \APACyear {2024}}%
}]{%
kouhen2024convective}
\APACinsertmetastar {%
kouhen2024convective}%
\begin{APACrefauthors}%
Kouhen, S.%
, Storer, B\BPBI A.%
, Aluie, H.%
, Marshall, D\BPBI P.%
\BCBL {}\ \BBA {} Christensen, H\BPBI M.%
\end{APACrefauthors}%
\unskip\
\newblock
\APACrefYearMonthDay{2024}{}{}.
\newblock
{\BBOQ}\APACrefatitle {Convective and orographic origins of the mesoscale
  kinetic energy spectrum} {Convective and orographic origins of the mesoscale
  kinetic energy spectrum}.{\BBCQ}
\newblock
\APACjournalVolNumPages{Geophysical Research Letters}{51}{21}{e2024GL110804}.
\PrintBackRefs{\CurrentBib}

\bibitem [\protect \citeauthoryear {%
{Leonard}%
}{%
{Leonard}%
}{%
{\protect \APACyear {1974}}%
}]{%
Leonard74}
\APACinsertmetastar {%
Leonard74}%
\begin{APACrefauthors}%
{Leonard}, A.%
\end{APACrefauthors}%
\unskip\
\newblock
\APACrefYearMonthDay{1974}{}{}.
\newblock
{\BBOQ}\APACrefatitle {{Energy Cascade in Large-Eddy Simulations of Turbulent
  Fluid Flows}} {{Energy Cascade in Large-Eddy Simulations of Turbulent Fluid
  Flows}}.{\BBCQ}
\newblock
\APACjournalVolNumPages{Adv. Geophys.}{18}{}{A237}.
\PrintBackRefs{\CurrentBib}

\bibitem [\protect \citeauthoryear {%
Li%
\ \protect \BOthers {.}}{%
Li%
\ \protect \BOthers {.}}{%
{\protect \APACyear {2024}}%
}]{%
li2024eddy}
\APACinsertmetastar {%
li2024eddy}%
\begin{APACrefauthors}%
Li, X.%
, Wang, Q.%
, Danilov, S.%
, Koldunov, N.%
, Liu, C.%
, M{\"u}ller, V.%
\BDBL {}Jung, T.%
\end{APACrefauthors}%
\unskip\
\newblock
\APACrefYearMonthDay{2024}{}{}.
\newblock
{\BBOQ}\APACrefatitle {Eddy activity in the Arctic Ocean projected to surge in
  a warming world} {Eddy activity in the arctic ocean projected to surge in a
  warming world}.{\BBCQ}
\newblock
\APACjournalVolNumPages{Nature Climate Change}{14}{2}{156--162}.
\PrintBackRefs{\CurrentBib}

\bibitem [\protect \citeauthoryear {%
Liu%
\ \protect \BOthers {.}}{%
Liu%
\ \protect \BOthers {.}}{%
{\protect \APACyear {2024}}%
}]{%
liu2024spatial}
\APACinsertmetastar {%
liu2024spatial}%
\begin{APACrefauthors}%
Liu, C.%
, Wang, Q.%
, Danilov, S.%
, Koldunov, N.%
, M{\"u}ller, V.%
, Li, X.%
\BDBL {}Zhang, S.%
\end{APACrefauthors}%
\unskip\
\newblock
\APACrefYearMonthDay{2024}{}{}.
\newblock
{\BBOQ}\APACrefatitle {Spatial scales of kinetic energy in the Arctic Ocean}
  {Spatial scales of kinetic energy in the arctic ocean}.{\BBCQ}
\newblock
\APACjournalVolNumPages{Journal of Geophysical Research:
  Oceans}{129}{3}{e2023JC020013}.
\PrintBackRefs{\CurrentBib}

\bibitem [\protect \citeauthoryear {%
Loose%
\ \protect \BOthers {.}}{%
Loose%
\ \protect \BOthers {.}}{%
{\protect \APACyear {2023}}%
}]{%
loose2023comparing}
\APACinsertmetastar {%
loose2023comparing}%
\begin{APACrefauthors}%
Loose, N.%
, Marques, G\BPBI M.%
, Adcroft, A.%
, Bachman, S.%
, Griffies, S\BPBI M.%
, Grooms, I.%
\BDBL {}Jansen, M\BPBI F.%
\end{APACrefauthors}%
\unskip\
\newblock
\APACrefYearMonthDay{2023}{}{}.
\newblock
{\BBOQ}\APACrefatitle {Comparing two parameterizations for the restratification
  effect of mesoscale eddies in an isopycnal ocean model} {Comparing two
  parameterizations for the restratification effect of mesoscale eddies in an
  isopycnal ocean model}.{\BBCQ}
\newblock
\APACjournalVolNumPages{Journal of Advances in Modeling Earth
  Systems}{15}{12}{e2022MS003518}.
\PrintBackRefs{\CurrentBib}

\bibitem [\protect \citeauthoryear {%
Meneveau%
}{%
Meneveau%
}{%
{\protect \APACyear {1994}}%
}]{%
Meneveau1994}
\APACinsertmetastar {%
Meneveau1994}%
\begin{APACrefauthors}%
Meneveau, C.%
\end{APACrefauthors}%
\unskip\
\newblock
\APACrefYearMonthDay{1994}{{\APACmonth{02}}}{}.
\newblock
{\BBOQ}\APACrefatitle {{Statistics of Turbulence Subgrid-Scale Stresses -
  Necessary Conditions and Experimental Tests}} {{Statistics of Turbulence
  Subgrid-Scale Stresses - Necessary Conditions and Experimental
  Tests}}.{\BBCQ}
\newblock
\APACjournalVolNumPages{Physics of Fluids}{6}{2}{815--833}.
\PrintBackRefs{\CurrentBib}

\bibitem [\protect \citeauthoryear {%
Mittal%
\ \BBA {} Iaccarino%
}{%
Mittal%
\ \BBA {} Iaccarino%
}{%
{\protect \APACyear {2005}}%
}]{%
mittal2005immersed}
\APACinsertmetastar {%
mittal2005immersed}%
\begin{APACrefauthors}%
Mittal, R.%
\BCBT {}\ \BBA {} Iaccarino, G.%
\end{APACrefauthors}%
\unskip\
\newblock
\APACrefYearMonthDay{2005}{}{}.
\newblock
{\BBOQ}\APACrefatitle {Immersed boundary methods} {Immersed boundary
  methods}.{\BBCQ}
\newblock
\APACjournalVolNumPages{Annu. Rev. Fluid Mech.}{37}{1}{239--261}.
\PrintBackRefs{\CurrentBib}

\bibitem [\protect \citeauthoryear {%
Peskin%
}{%
Peskin%
}{%
{\protect \APACyear {2002}}%
}]{%
peskin2002immersed}
\APACinsertmetastar {%
peskin2002immersed}%
\begin{APACrefauthors}%
Peskin, C\BPBI S.%
\end{APACrefauthors}%
\unskip\
\newblock
\APACrefYearMonthDay{2002}{}{}.
\newblock
{\BBOQ}\APACrefatitle {The immersed boundary method} {The immersed boundary
  method}.{\BBCQ}
\newblock
\APACjournalVolNumPages{Acta numerica}{11}{}{479--517}.
\PrintBackRefs{\CurrentBib}

\bibitem [\protect \citeauthoryear {%
Press%
}{%
Press%
}{%
{\protect \APACyear {2007}}%
}]{%
press2007numerical}
\APACinsertmetastar {%
press2007numerical}%
\begin{APACrefauthors}%
Press, W\BPBI H.%
\end{APACrefauthors}%
\unskip\
\newblock
\APACrefYear{2007}.
\newblock
\APACrefbtitle {Numerical recipes 3rd edition: The art of scientific computing}
  {Numerical recipes 3rd edition: The art of scientific computing}.
\newblock
\APACaddressPublisher{}{Cambridge university press}.
\PrintBackRefs{\CurrentBib}

\bibitem [\protect \citeauthoryear {%
Rai%
, Farrar%
\BCBL {}\ \BBA {} Aluie%
}{%
Rai%
\ \protect \BOthers {.}}{%
{\protect \APACyear {2025}}%
}]{%
rai2025atmospheric}
\APACinsertmetastar {%
rai2025atmospheric}%
\begin{APACrefauthors}%
Rai, S.%
, Farrar, J\BPBI T.%
\BCBL {}\ \BBA {} Aluie, H.%
\end{APACrefauthors}%
\unskip\
\newblock
\APACrefYearMonthDay{2025}{}{}.
\newblock
{\BBOQ}\APACrefatitle {Atmospheric wind energization of ocean weather}
  {Atmospheric wind energization of ocean weather}.{\BBCQ}
\newblock
\APACjournalVolNumPages{Nature Communications}{16}{1}{1172}.
\PrintBackRefs{\CurrentBib}

\bibitem [\protect \citeauthoryear {%
Rai%
, Hecht%
, Maltrud%
\BCBL {}\ \BBA {} Aluie%
}{%
Rai%
\ \protect \BOthers {.}}{%
{\protect \APACyear {2021}}%
}]{%
rai2021scale}
\APACinsertmetastar {%
rai2021scale}%
\begin{APACrefauthors}%
Rai, S.%
, Hecht, M.%
, Maltrud, M.%
\BCBL {}\ \BBA {} Aluie, H.%
\end{APACrefauthors}%
\unskip\
\newblock
\APACrefYearMonthDay{2021}{}{}.
\newblock
{\BBOQ}\APACrefatitle {Scale of oceanic eddy killing by wind from global
  satellite observations} {Scale of oceanic eddy killing by wind from global
  satellite observations}.{\BBCQ}
\newblock
\APACjournalVolNumPages{Science Advances}{7}{28}{eabf4920}.
\PrintBackRefs{\CurrentBib}

\bibitem [\protect \citeauthoryear {%
Rayleigh%
}{%
Rayleigh%
}{%
{\protect \APACyear {1883}}%
}]{%
Rayleigh83}
\APACinsertmetastar {%
Rayleigh83}%
\begin{APACrefauthors}%
Rayleigh.%
\end{APACrefauthors}%
\unskip\
\newblock
\APACrefYearMonthDay{1883}{}{}.
\newblock
{\BBOQ}\APACrefatitle {Investigation of the Character of the Equilibrium of an
  Incompressible Heavy Fluid of Variable Density} {Investigation of the
  character of the equilibrium of an incompressible heavy fluid of variable
  density}.{\BBCQ}
\newblock
\APACjournalVolNumPages{Proceedings of the London Mathematical
  Society}{s1-14}{1}{170--177}.
\newblock
\begin{APACrefURL} \url{http://dx.doi.org/10.1112/plms/s1-14.1.170}
  \end{APACrefURL}
\newblock
\begin{APACrefDOI} \doi{10.1112/plms/s1-14.1.170} \end{APACrefDOI}
\PrintBackRefs{\CurrentBib}

\bibitem [\protect \citeauthoryear {%
Schubert%
, Gula%
, Greatbatch%
, Baschek%
\BCBL {}\ \BBA {} Biastoch%
}{%
Schubert%
\ \protect \BOthers {.}}{%
{\protect \APACyear {2020}}%
}]{%
schubert2020submesoscale}
\APACinsertmetastar {%
schubert2020submesoscale}%
\begin{APACrefauthors}%
Schubert, R.%
, Gula, J.%
, Greatbatch, R\BPBI J.%
, Baschek, B.%
\BCBL {}\ \BBA {} Biastoch, A.%
\end{APACrefauthors}%
\unskip\
\newblock
\APACrefYearMonthDay{2020}{}{}.
\newblock
{\BBOQ}\APACrefatitle {The submesoscale kinetic energy cascade: Mesoscale
  absorption of submesoscale mixed layer eddies and frontal downscale fluxes}
  {The submesoscale kinetic energy cascade: Mesoscale absorption of
  submesoscale mixed layer eddies and frontal downscale fluxes}.{\BBCQ}
\newblock
\APACjournalVolNumPages{Journal of Physical Oceanography}{50}{9}{2573--2589}.
\PrintBackRefs{\CurrentBib}

\bibitem [\protect \citeauthoryear {%
Schubert%
, Vergara%
\BCBL {}\ \BBA {} Gula%
}{%
Schubert%
\ \protect \BOthers {.}}{%
{\protect \APACyear {2023}}%
}]{%
schubert2023open}
\APACinsertmetastar {%
schubert2023open}%
\begin{APACrefauthors}%
Schubert, R.%
, Vergara, O.%
\BCBL {}\ \BBA {} Gula, J.%
\end{APACrefauthors}%
\unskip\
\newblock
\APACrefYearMonthDay{2023}{}{}.
\newblock
{\BBOQ}\APACrefatitle {The open ocean kinetic energy cascade is strongest in
  late winter and spring} {The open ocean kinetic energy cascade is strongest
  in late winter and spring}.{\BBCQ}
\newblock
\APACjournalVolNumPages{Communications Earth \& Environment}{4}{1}{450}.
\PrintBackRefs{\CurrentBib}

\bibitem [\protect \citeauthoryear {%
Shaham%
\ \BBA {} Barkan%
}{%
Shaham%
\ \BBA {} Barkan%
}{%
{\protect \APACyear {2025}}%
}]{%
shaham2025spectral}
\APACinsertmetastar {%
shaham2025spectral}%
\begin{APACrefauthors}%
Shaham, M.%
\BCBT {}\ \BBA {} Barkan, R.%
\end{APACrefauthors}%
\unskip\
\newblock
\APACrefYearMonthDay{2025}{}{}.
\newblock
{\BBOQ}\APACrefatitle {Spectral flux decomposition in a wind-driven channel
  flow with near-inertial waves} {Spectral flux decomposition in a wind-driven
  channel flow with near-inertial waves}.{\BBCQ}
\newblock
\APACjournalVolNumPages{Journal of Advances in Modeling Earth
  Systems}{17}{1}{e2023MS004036}.
\PrintBackRefs{\CurrentBib}

\bibitem [\protect \citeauthoryear {%
Solodoch%
\ \protect \BOthers {.}}{%
Solodoch%
\ \protect \BOthers {.}}{%
{\protect \APACyear {2023}}%
}]{%
solodoch2023basin}
\APACinsertmetastar {%
solodoch2023basin}%
\begin{APACrefauthors}%
Solodoch, A.%
, Barkan, R.%
, Verma, V.%
, Gildor, H.%
, Toledo, Y.%
, Khain, P.%
\BCBL {}\ \BBA {} Levi, Y.%
\end{APACrefauthors}%
\unskip\
\newblock
\APACrefYearMonthDay{2023}{}{}.
\newblock
{\BBOQ}\APACrefatitle {Basin-Scale to Submesoscale Variability of the East
  Mediterranean Sea Upper Circulation} {Basin-scale to submesoscale variability
  of the east mediterranean sea upper circulation}.{\BBCQ}
\newblock
\APACjournalVolNumPages{Journal of Physical Oceanography}{53}{9}{2137--2158}.
\PrintBackRefs{\CurrentBib}

\bibitem [\protect \citeauthoryear {%
Soltani~Tehrani%
\ \BBA {} Aluie%
}{%
Soltani~Tehrani%
\ \BBA {} Aluie%
}{%
{\protect \APACyear {2023}}%
}]{%
soltani2023galilean}
\APACinsertmetastar {%
soltani2023galilean}%
\begin{APACrefauthors}%
Soltani~Tehrani, D.%
\BCBT {}\ \BBA {} Aluie, H.%
\end{APACrefauthors}%
\unskip\
\newblock
\APACrefYearMonthDay{2023}{}{}.
\newblock
{\BBOQ}\APACrefatitle {On Galilean invariance of mean kinetic helicity} {On
  galilean invariance of mean kinetic helicity}.{\BBCQ}
\newblock
\APACjournalVolNumPages{Physics of Fluids}{35}{12}{121701}.
\PrintBackRefs{\CurrentBib}

\bibitem [\protect \citeauthoryear {%
Srinivasan%
, McWilliams%
, Molemaker%
\BCBL {}\ \BBA {} Barkan%
}{%
Srinivasan%
\ \protect \BOthers {.}}{%
{\protect \APACyear {2019}}%
}]{%
srinivasan2019submesoscale}
\APACinsertmetastar {%
srinivasan2019submesoscale}%
\begin{APACrefauthors}%
Srinivasan, K.%
, McWilliams, J\BPBI C.%
, Molemaker, M\BPBI J.%
\BCBL {}\ \BBA {} Barkan, R.%
\end{APACrefauthors}%
\unskip\
\newblock
\APACrefYearMonthDay{2019}{}{}.
\newblock
{\BBOQ}\APACrefatitle {Submesoscale vortical wakes in the lee of topography}
  {Submesoscale vortical wakes in the lee of topography}.{\BBCQ}
\newblock
\APACjournalVolNumPages{Journal of Physical Oceanography}{49}{7}{1949--1971}.
\PrintBackRefs{\CurrentBib}

\bibitem [\protect \citeauthoryear {%
Steinberg%
, Cole%
, Drushka%
\BCBL {}\ \BBA {} Abernathey%
}{%
Steinberg%
\ \protect \BOthers {.}}{%
{\protect \APACyear {2022}}%
}]{%
steinberg2022seasonality}
\APACinsertmetastar {%
steinberg2022seasonality}%
\begin{APACrefauthors}%
Steinberg, J\BPBI M.%
, Cole, S\BPBI T.%
, Drushka, K.%
\BCBL {}\ \BBA {} Abernathey, R\BPBI P.%
\end{APACrefauthors}%
\unskip\
\newblock
\APACrefYearMonthDay{2022}{}{}.
\newblock
{\BBOQ}\APACrefatitle {Seasonality of the mesoscale inverse cascade as inferred
  from global scale-dependent eddy energy observations} {Seasonality of the
  mesoscale inverse cascade as inferred from global scale-dependent eddy energy
  observations}.{\BBCQ}
\newblock
\APACjournalVolNumPages{Journal of Physical Oceanography}{52}{8}{1677--1691}.
\PrintBackRefs{\CurrentBib}

\bibitem [\protect \citeauthoryear {%
Storer%
, Buzzicotti%
, Khatri%
, Griffies%
\BCBL {}\ \BBA {} Aluie%
}{%
Storer%
\ \protect \BOthers {.}}{%
{\protect \APACyear {2022}}%
}]{%
storer2022global}
\APACinsertmetastar {%
storer2022global}%
\begin{APACrefauthors}%
Storer, B\BPBI A.%
, Buzzicotti, M.%
, Khatri, H.%
, Griffies, S\BPBI M.%
\BCBL {}\ \BBA {} Aluie, H.%
\end{APACrefauthors}%
\unskip\
\newblock
\APACrefYearMonthDay{2022}{}{}.
\newblock
{\BBOQ}\APACrefatitle {Global energy spectrum of the general oceanic
  circulation} {Global energy spectrum of the general oceanic
  circulation}.{\BBCQ}
\newblock
\APACjournalVolNumPages{Nature communications}{13}{1}{5314}.
\PrintBackRefs{\CurrentBib}

\bibitem [\protect \citeauthoryear {%
Storer%
, Buzzicotti%
, Khatri%
, Griffies%
\BCBL {}\ \BBA {} Aluie%
}{%
Storer%
\ \protect \BOthers {.}}{%
{\protect \APACyear {2023}}%
}]{%
storer2023global}
\APACinsertmetastar {%
storer2023global}%
\begin{APACrefauthors}%
Storer, B\BPBI A.%
, Buzzicotti, M.%
, Khatri, H.%
, Griffies, S\BPBI M.%
\BCBL {}\ \BBA {} Aluie, H.%
\end{APACrefauthors}%
\unskip\
\newblock
\APACrefYearMonthDay{2023}{}{}.
\newblock
{\BBOQ}\APACrefatitle {Global cascade of kinetic energy in the ocean and the
  atmospheric imprint} {Global cascade of kinetic energy in the ocean and the
  atmospheric imprint}.{\BBCQ}
\newblock
\APACjournalVolNumPages{Science advances}{9}{51}{eadi7420}.
\PrintBackRefs{\CurrentBib}

\bibitem [\protect \citeauthoryear {%
Taylor%
}{%
Taylor%
}{%
{\protect \APACyear {1950}}%
}]{%
Taylor50}
\APACinsertmetastar {%
Taylor50}%
\begin{APACrefauthors}%
Taylor, G.%
\end{APACrefauthors}%
\unskip\
\newblock
\APACrefYearMonthDay{1950}{}{}.
\newblock
{\BBOQ}\APACrefatitle {The instability of liquid surfaces when accelerated in a
  direction perpendicular to their planes. I} {The instability of liquid
  surfaces when accelerated in a direction perpendicular to their planes.
  i}.{\BBCQ}
\newblock
\APACjournalVolNumPages{Proceedings of the Royal Society of London. Series A.
  Mathematical and Physical Sciences}{201}{1065}{192--196}.
\PrintBackRefs{\CurrentBib}

\bibitem [\protect \citeauthoryear {%
Trottenberg%
, Oosterlee%
\BCBL {}\ \BBA {} Schuller%
}{%
Trottenberg%
\ \protect \BOthers {.}}{%
{\protect \APACyear {2001}}%
}]{%
trottenberg2001multigrid}
\APACinsertmetastar {%
trottenberg2001multigrid}%
\begin{APACrefauthors}%
Trottenberg, U.%
, Oosterlee, C\BPBI W.%
\BCBL {}\ \BBA {} Schuller, A.%
\end{APACrefauthors}%
\unskip\
\newblock
\APACrefYear{2001}.
\newblock
\APACrefbtitle {Multigrid methods} {Multigrid methods}.
\newblock
\APACaddressPublisher{}{Academic press}.
\PrintBackRefs{\CurrentBib}

\bibitem [\protect \citeauthoryear {%
Wunsch%
}{%
Wunsch%
}{%
{\protect \APACyear {1991}}%
}]{%
wunsch1991global}
\APACinsertmetastar {%
wunsch1991global}%
\begin{APACrefauthors}%
Wunsch, C.%
\end{APACrefauthors}%
\unskip\
\newblock
\APACrefYearMonthDay{1991}{}{}.
\newblock
{\BBOQ}\APACrefatitle {Global-scale sea surface variability from combined
  altimetric and tide gauge measurements} {Global-scale sea surface variability
  from combined altimetric and tide gauge measurements}.{\BBCQ}
\newblock
\APACjournalVolNumPages{Journal of Geophysical Research:
  Oceans}{96}{C8}{15053--15082}.
\PrintBackRefs{\CurrentBib}

\bibitem [\protect \citeauthoryear {%
Wunsch%
\ \BBA {} Stammer%
}{%
Wunsch%
\ \BBA {} Stammer%
}{%
{\protect \APACyear {1995}}%
}]{%
wunsch1995global}
\APACinsertmetastar {%
wunsch1995global}%
\begin{APACrefauthors}%
Wunsch, C.%
\BCBT {}\ \BBA {} Stammer, D.%
\end{APACrefauthors}%
\unskip\
\newblock
\APACrefYearMonthDay{1995}{}{}.
\newblock
{\BBOQ}\APACrefatitle {The global frequency-wavenumber spectrum of oceanic
  variability estimated from TOPEX/POSEIDON altimetric measurements} {The
  global frequency-wavenumber spectrum of oceanic variability estimated from
  topex/poseidon altimetric measurements}.{\BBCQ}
\newblock
\APACjournalVolNumPages{Journal of Geophysical Research:
  Oceans}{100}{C12}{24895--24910}.
\PrintBackRefs{\CurrentBib}

\bibitem [\protect \citeauthoryear {%
Xue%
, Storer%
, Glade%
\BCBL {}\ \BBA {} Aluie%
}{%
Xue%
\ \protect \BOthers {.}}{%
{\protect \APACyear {2024}}%
}]{%
xue2024surface}
\APACinsertmetastar {%
xue2024surface}%
\begin{APACrefauthors}%
Xue, S.%
, Storer, B\BPBI A.%
, Glade, R\BPBI C.%
\BCBL {}\ \BBA {} Aluie, H.%
\end{APACrefauthors}%
\unskip\
\newblock
\APACrefYearMonthDay{2024}{}{}.
\newblock
{\BBOQ}\APACrefatitle {Surface variability mapping and roughness analysis of
  the moon using a coarse-graining decomposition} {Surface variability mapping
  and roughness analysis of the moon using a coarse-graining
  decomposition}.{\BBCQ}
\newblock
\APACjournalVolNumPages{Journal of Geophysical Research:
  Planets}{129}{10}{e2024JE008484}.
\PrintBackRefs{\CurrentBib}

\bibitem [\protect \citeauthoryear {%
Yu%
, Barkan%
\BCBL {}\ \BBA {} Naveira~Garabato%
}{%
Yu%
\ \protect \BOthers {.}}{%
{\protect \APACyear {2024}}%
}]{%
yu2024intensification}
\APACinsertmetastar {%
yu2024intensification}%
\begin{APACrefauthors}%
Yu, X.%
, Barkan, R.%
\BCBL {}\ \BBA {} Naveira~Garabato, A\BPBI C.%
\end{APACrefauthors}%
\unskip\
\newblock
\APACrefYearMonthDay{2024}{}{}.
\newblock
{\BBOQ}\APACrefatitle {Intensification of submesoscale frontogenesis and
  forward energy cascade driven by upper-ocean convergent flows}
  {Intensification of submesoscale frontogenesis and forward energy cascade
  driven by upper-ocean convergent flows}.{\BBCQ}
\newblock
\APACjournalVolNumPages{Nature Communications}{15}{1}{9214}.
\PrintBackRefs{\CurrentBib}

\bibitem [\protect \citeauthoryear {%
Zhang%
, Perezhogin%
, Adcroft%
\BCBL {}\ \BBA {} Zanna%
}{%
Zhang%
\ \protect \BOthers {.}}{%
{\protect \APACyear {2025}}%
}]{%
zhang2025addressing}
\APACinsertmetastar {%
zhang2025addressing}%
\begin{APACrefauthors}%
Zhang, C.%
, Perezhogin, P.%
, Adcroft, A.%
\BCBL {}\ \BBA {} Zanna, L.%
\end{APACrefauthors}%
\unskip\
\newblock
\APACrefYearMonthDay{2025}{}{}.
\newblock
{\BBOQ}\APACrefatitle {Addressing out-of-sample issues in multi-layer
  convolutional neural-network parameterization of mesoscale eddies applied
  near coastlines} {Addressing out-of-sample issues in multi-layer
  convolutional neural-network parameterization of mesoscale eddies applied
  near coastlines}.{\BBCQ}
\newblock
\APACjournalVolNumPages{Journal of Advances in Modeling Earth
  Systems}{17}{5}{e2024MS004819}.
\PrintBackRefs{\CurrentBib}

\bibitem [\protect \citeauthoryear {%
Zhao%
\ \BBA {} Aluie%
}{%
Zhao%
\ \BBA {} Aluie%
}{%
{\protect \APACyear {2018}}%
}]{%
zhao2018inviscid}
\APACinsertmetastar {%
zhao2018inviscid}%
\begin{APACrefauthors}%
Zhao, D.%
\BCBT {}\ \BBA {} Aluie, H.%
\end{APACrefauthors}%
\unskip\
\newblock
\APACrefYearMonthDay{2018}{}{}.
\newblock
{\BBOQ}\APACrefatitle {Inviscid criterion for decomposing scales} {Inviscid
  criterion for decomposing scales}.{\BBCQ}
\newblock
\APACjournalVolNumPages{Physical Review Fluids}{3}{5}{054603}.
\PrintBackRefs{\CurrentBib}

\bibitem [\protect \citeauthoryear {%
Zhao%
, Aluie%
\BCBL {}\ \BBA {} Li%
}{%
Zhao%
\ \protect \BOthers {.}}{%
{\protect \APACyear {2025}}%
}]{%
zhao2025multi}
\APACinsertmetastar {%
zhao2025multi}%
\begin{APACrefauthors}%
Zhao, D.%
, Aluie, H.%
\BCBL {}\ \BBA {} Li, G.%
\end{APACrefauthors}%
\unskip\
\newblock
\APACrefYearMonthDay{2025}{}{}.
\newblock
{\BBOQ}\APACrefatitle {Multi-scale dynamics of scalar transfer in
  Rayleigh--Taylor turbulent mixing} {Multi-scale dynamics of scalar transfer
  in rayleigh--taylor turbulent mixing}.{\BBCQ}
\newblock
\APACjournalVolNumPages{Journal of Fluid Mechanics}{1011}{}{A4}.
\PrintBackRefs{\CurrentBib}

\bibitem [\protect \citeauthoryear {%
Zhao%
, Betti%
\BCBL {}\ \BBA {} Aluie%
}{%
Zhao%
\ \protect \BOthers {.}}{%
{\protect \APACyear {2022}}%
}]{%
zhao2022scale}
\APACinsertmetastar {%
zhao2022scale}%
\begin{APACrefauthors}%
Zhao, D.%
, Betti, R.%
\BCBL {}\ \BBA {} Aluie, H.%
\end{APACrefauthors}%
\unskip\
\newblock
\APACrefYearMonthDay{2022}{}{}.
\newblock
{\BBOQ}\APACrefatitle {Scale interactions and anisotropy in Rayleigh--Taylor
  turbulence} {Scale interactions and anisotropy in rayleigh--taylor
  turbulence}.{\BBCQ}
\newblock
\APACjournalVolNumPages{Journal of Fluid Mechanics}{930}{}{A29}.
\PrintBackRefs{\CurrentBib}

\end{thebibliography}
\end{document}